\documentclass[12pt,a4paper]{article}
\usepackage[latin1]{inputenc}
\usepackage{amsmath}
\usepackage{amsfonts}
\usepackage{amssymb}
\usepackage{graphicx,a4wide}
\usepackage{verbatim,cite}

\usepackage{color}
\usepackage{mathrsfs}
\DeclareMathAlphabet{\mathpzc}{OT1}{pzc}{m}{it}

\usepackage{amsmath}
\usepackage{setspace}

\usepackage[mathscr]{euscript}

\usepackage[hidelinks]{hyperref}

\def\cO{{\mathcal O}}
\def\cN{{\mathcal N}}
\def\cP{{\mathcal P}}
\def\ula{{\underline{\smash \lambda}}}

\def\symbol	 {\mathscr{M}}
\def\symbolB     {B}
\def\LL		{{\mathbb L}}

\def\Db		{\overline{D}}
\def\Dbar		{\overline{D}}

\def\cH		{\mathcal{H}}
\def\norma	{\mathscr{N}}
\def\cK		{\mathcal{K}}
\def\du 		{\partial_u}
\def\dv 		{\partial_v}
\def\a		{\mathsf{a}}
\def\R		{\mathsf{R}}
\def\cA		{\mathcal{A}}
\def\d		{g}

\def\CB		{\mathcal{B}}
\def\bF		{{\bf F}}
\def\bP		{{\bf P}}

\newcommand{\beq}{\begin{equation}}
\newcommand{\eeq}{\end{equation}}
\newcommand{\bea}{\begin{eqnarray}}
\newcommand{\eea}{\end{eqnarray}}

\def \be  {\begin{equation}}
\def \ee  {\end{equation}}
\def \ba  {\begin{eqnarray}}
\def \ea  {\end{eqnarray}}

\begin{document}

\thispagestyle{empty}

\null\vskip-43pt \hfill
\begin{minipage}[t]{30mm}
	DCPT-17/33
\end{minipage}

\null\vskip-12pt \hfill  \\
\null\vskip-12pt \hfill   \\

\vskip2.2truecm
\begin{center}
	\vskip 0.2truecm {\Large\bf
		{\Large Loop corrections for Kaluza-Klein AdS amplitudes}
	}\\
	\vskip 1truecm
	{\bf F.~Aprile${}^{1}$, J.~M. Drummond${}^{2}$, P.~Heslop${}^{3}$, H.~Paul${}^{2}$ \\
	}
	
	\vskip 0.4truecm
	
	{\it
		${}^{1}$ Dipartimento di Fisica, Universit\`a di Milano-Bicocca \& INFN, 
		Sezione di Milano-Bicocca, I-20126 Milano,\\
		\vskip .2truecm }
	\vskip .2truecm
	{\it
		${}^{2}$ School of Physics and Astronomy and STAG Research Centre, \\
		University of Southampton,
		Highfield,  SO17 1BJ,\\
		\vskip .2truecm                        }
	\vskip .2truecm
	{\it
		${}^{3}$ Mathematics Department, Durham University, \\
		Science Laboratories, South Rd, Durham DH1 3LE \vskip .2truecm                        }
\end{center}

\vskip 1truecm 
\centerline{\bf Abstract} 

Recently we conjectured the four-point amplitude of graviton multiplets in ${\rm AdS}_5 \times {\rm S}^5$ at one loop by exploiting the operator product expansion of $\cN=4$ super Yang-Mills theory. Here we give the first extension of those results to include Kaluza-Klein modes, obtaining the amplitude for two graviton multiplets and two states of the first KK mode. Our method again relies on resolving the large N degeneracy among a family of long double-trace operators, for which we obtain explicit formulas for the leading anomalous dimensions.  Having constructed the one-loop amplitude we are able to obtain a formula for the one-loop corrections to the anomalous dimensions of all twist five double-trace operators.

\medskip

\noindent

\newpage
\setcounter{page}{1}\setcounter{footnote}{0}


\section{Introduction and summary of results}

The AdS/CFT correspondence \cite{1,2,3} relates correlation functions of the boundary CFT to on-shell AdS amplitudes of bulk fields. While there has been much study of tree-level bulk processes \cite{Liu:1998ty,Arutyunov:1999fb,Arutyunov:2002fh,Arutyunov:2003ae,Dolan:2006ec,Berdichevsky:2007xd,Uruchurtu:2008kp,Uruchurtu:2011wh,Rastelli:2016nze}, until recently there had not been much work on loop corrections (see~\cite{Cardona:2017tsw,Giombi:2017hpr} for some recent work in this direction). Essentially this is because such loop computations are extremely challenging from the bulk approach. Recently however, another approach based on the OPE structure of the boundary CFT has been initiated \cite{1612.03891,Alday:2017xua,Aprile:2017bgs}. In \cite{Aprile:2017bgs} we used the OPE structure of $\mathcal{N}=4$ super Yang-Mills theory to bootstrap the one-loop correction to the AdS${}_5$ scattering of four-graviton multiplets, or equivalently the $1/N^4$ correction to the four-point correlator of four energy-momentum multiplets in the large $N$ limit.

A crucial ingredient in the analysis is the resolution of a large degeneracy among the spectrum of double-trace operators which occurs in the strict large $N$ limit. The relevant explicit formulas for the anomalous dimensions and three-point functions of double-trace operators were obtained by considering multiple correlators which exhibit the same exchanged operators in their OPE decompositions \cite{unmixing}. Such data is available due to a remarkably compact formula \cite{Rastelli:2016nze} for all four-point tree-level scattering processes of graviton multiplets or their associated Kaluza-Klein modes which are present in the five-dimensional bulk due to the reduction from ten dimensions on $S^5$. 

Here we will summarise the results obtained in \cite{Aprile:2017bgs,unmixing} which allowed for the bootstrapping of the one-loop four-graviton amplitude. Firstly we may consider the following set of double-trace operators $K_{t,l,n,i}$, labelled by $i$ which runs from $1$ to $(t-n-1)$,
\be
\{ \mathcal{O}_{n+2} \Box^{t-n-2} \partial^l \mathcal{O}_{n+2}, \mathcal{O}_{n+3} \Box^{t-n-3} \partial^l \mathcal{O}_{n+3}, \ldots, \mathcal{O}_{t} \Box^0 \partial^l \mathcal{O}_{t}\}\Big|_{[n,0,n]}\,.
\ee
At large $N$ the above operators are degenerate; they all have large $N$ dimensions $\Delta=2t+l$, spin $l$ and $SU(4)$ labels $[n,0,n]$. However, by considering the correlation functions of the form $\langle \mathcal{O}_p \mathcal{O}_p \mathcal{O}_q \mathcal{O}_q \rangle$, we can extract the order $1/N^2$ anomalous dimensions and resolve the degeneracy. The large $N$ expansion of the dimensions takes the form
\be
\Delta = 2t + l + \frac{2}{N^2} \eta^{(1)}_{n,t,l,i} + O(1/N^4)\,,
\ee
with the anomalous dimensions given by
\be
\label{anomn0n}
\eta^{(1)}_{t,l,n,i} = -\frac{2(t-n-1)(t)_2(t+n+2)(t+l-n)(t+l+1)_2(t+l+n+3)}{(l+2i+n-1)_6}\,,
\ee
where we used the notation $(x)_n = x(x+1)\ldots(x+n-1)$ for the Pochhammer symbol.

To bootstrap the one-loop graviton amplitude the dimensions in the singlet channel (i.e. the case $n=0$ above) were needed together with the results for the leading order three-point functions $\langle \mathcal{O}_2 \mathcal{O}_2 K_{t,l,0,i} \rangle$, for which we also gave a closed form expression in \cite{Aprile:2017bgs,unmixing},
\be
\langle \mathcal{O}_{2} \mathcal{O}_{2} K_{t,l,0,i} \rangle^2 = \frac{8(t+l+1)!^2 t!^2 (l+1)(2t+l+2)}{(2t)!(2t+2l+2)!}\, \R_{t,l,i}\, \a_{t,i}
\ee
where
\begin{align}
\R_{t,l,i} &= \frac{2^{1-t}(2l+3+4i) (l+i+1)_{t-i-1} (t+l+4)_{i-1}}{(\tfrac{5}{2}+l+i)_{t-1}}\,,\notag \\
\a_{t,i} &= \frac{2^{(1 - t)} (2 + 2 i)! (t-2)! (2t - 2 i + 2)!}{3 (i-1)! (i+1)! (t+2)! (t-i-1)! (t-i+1)!}\,.\label{3pnt}
\end{align}

These results allowed us to predict the double discontinuity of the the correlator at order $1/N^4$. We were then able to construct a crossing symmetric function with the correct double discontinuities. Remarkably the function we obtained was expressed in terms of the four-dimensional one-loop and two-loop ladder integrals. Having obtained such a result we were then able to extract from it a closed form for all spins for the next correction to the anomalous dimensions for the twist-four singlet operators (expanding in $a = 1/(N^2-1)$),
\be
\Delta = 4 + l + 2 a \eta_{l}^{(1)} + 2 a^2 \eta_l^{(2)} + O(a^3)\,,
\ee
where
\be
\label{twist4one-loopanomdims}
\eta_l^{(2)} = \frac{1344 (l-7) (l+14)}{(l-1) (l+1)^2 (l+6)^2 (l+8)} -\frac{2304 (2 l+7)}{(l+1)^3 (l+6)^3}\,.
\ee
The cases of the above formula for $l=2,4$ were also quoted in \cite{Alday:2017xua}.

In the present work we would like to generalise our analysis to include scattering of multiplets of Kaluza Klein states. The simplest case of such an amplitude is the one for the scattering of two graviton multiplets and two Kaluza Klein states of the next level. This corresponds to the CFT correlator $\langle \mathcal{O}_2 \mathcal{O}_2 \mathcal{O}_3 \mathcal{O}_3 \rangle$. This generalisation introduces a number of new challenges. Firstly we must deal with a correlator which has less crossing symmetry. This necessitates an OPE analysis in more than one crossing channel, or equivalently we must consider the crossed correlator $\langle \mathcal{O}_2  \mathcal{O}_3  \mathcal{O}_2  \mathcal{O}_3 \rangle$. Secondly, in pursuing such an OPE analysis, we must obtain information about operators with non-trivial $SU(4)$ representation labels, unlike the case of $\langle \mathcal{O}_2 \mathcal{O}_2 \mathcal{O}_2 \mathcal{O}_2\rangle$ where all exchanged long operators are singlets. We are then led to consider a new mixing problem involving correlators of the form $\langle \mathcal{O}_p \mathcal{O}_{p+1} \mathcal{O}_q \mathcal{O}_{q+1} \rangle$. Finally such correlators have both even and odd spin sectors in their OPE decomposition and we need to deal with both in order to construct the leading discontinuities for the one-loop amplitude. 

Once these technical obstacles are overcome we are able to proceed very similarly to the case of the $\langle \mathcal{O}_2 \mathcal{O}_2 \mathcal{O}_2 \mathcal{O}_2\rangle$ correlator and resolve the associated mixing problem. In particular we obtain an explicit result for the anomalous dimensions of the double trace operators $\mathcal{K}_{t,l,i}$ in the $[0,1,0]$ representation given by (for $i=1$ to $(t-1)$)
\be
\{\cK_{t,l,1},\cK_{t,l,2},\dots \cK_{t,l,t{-}1}\} \sim \{ \cO_2 \partial^l \Box^{t-2} \cO_3,\cO_3 \partial^l \Box^{t-3} \cO_{4}, \dots ,\cO_t \partial^l \cO_{t+1} \}|_{[0,1,0]}\,.
\ee
We find
\be
\Delta =  2t + l + 1 + 2 a \eta_{t,l,i}^{(1)} + O(a^2)\,,
\ee
with
\begin{align}
\eta^{(1)}_{t,l,i} = \left\{ \begin{array}{ll}
- \frac{2 \left(  t-1\right)_2 \left(  t+2\right)_2 \left(  l+t\right)_2\left(  l+t+3\right)_2 }{   \left( l+2i-1 \right)_{6} } \qquad &l=0,2,\dots\\[.2cm]
-\frac{2 \left(  t-1\right)_2 \left(  t+2\right)_2   \left(  l+t\right)_2\left(  l+t+3\right)_2 }{   \left( l+2i \right)_{6} } \qquad &\l=1,3,\dots
\end{array}\right.
\label{anom010}
\end{align}

With the solution to the mixing problem to hand we are able to explicitly construct the leading discontinuities in both channels.
We may then construct a function which reproduces the leading discontinuities in all channels. From this we can then extract new information about the dimensions of the twist-five operators in the $[0,1,0]$ representation analogously to (\ref{twist4one-loopanomdims})
\be
\Delta = 5 + l + 2 a \eta_{l}^{(1)} + 2 a^2 \eta_l^{(2)} + O(a^3)\,,
\ee
\begin{align}
\eta^{(2)}_{2,l} = \left\{ \begin{array}{ll}
\frac{  320 \left(9 l^4+68 l^3-1151 l^2-5738 l-3688\right)}{ (l-1) (l+1)^3 (l+4)^3 (l+8)} \qquad &l=2,4,\dots\\[.2cm]
\frac{ 320 \left(9 l^4+140 l^3-487 l^2-11262 l-29400\right)}{ l (l+4)^3 (l+7)^3 (l+9)}\qquad &\l=3,5,\dots
\end{array}\right.\label{anom0101loop}
\end{align}

{ The layout of the paper is as follows.}
In section~\ref{sec2} we review the general structure of four-point correlators of half-BPS multiplets, focussing on the cases relevant here, $\langle \mathcal{O}_2 \mathcal{O}_2 \mathcal{O}_3 \mathcal{O}_3 \rangle$ and $\langle \mathcal{O}_2 \mathcal{O}_3 \mathcal{O}_2 \mathcal{O}_3 \rangle$. Then in section~\ref{sec_Overview} we review the OPE and superblock expansion of these correlators, relevant for our purposes, highlighting the need to solve a mixing problem. In section~\ref{sec4} we go on to solve the mixing problem by analysing the correlators $\langle \mathcal{O}_p \mathcal{O}_p \mathcal{O}_q \mathcal{O}_q \rangle $ and $\langle \mathcal{O}_p \mathcal{O}_{p+1} \mathcal{O}_q \mathcal{O}_{q+1} \rangle $ at leading and next to leading order in large $N$. In section~\ref{sec5} we collect together this unmixed data to first derive the double discontinuity of the $1/N^4$  $\langle \mathcal{O}_2 \mathcal{O}_2 \mathcal{O}_3 \mathcal{O}_3 \rangle$ correlator in all channels, before uplifting the double discontinuity to the full result. In section~\ref{sec6} we use this uplifted function to derive new $O(1/N^4)$ anomalous dimensions for operators in the $[0,1,0]$ representation of $SU(4)$. In section~\ref{sec7} we point out a symmetry displayed by all the results derived here and previously at strong coupling. In two appendices we give details of the superblocks and the tree-level correlators used in the main body of the paper.

{\bf Note added:} During the preparation of this paper, \cite{Alday:2017vkk} appeared which also introduces the Casimir operator (\ref{defDelta8}) for the singlet channel in resumming the double discontinuity in the $\langle \mathcal{O}_2 \mathcal{O}_2 \mathcal{O}_2 \mathcal{O}_2 \rangle$ case.

\section{Four-point correlators of half-BPS operators}
\label{sec2}
The basic objects we wish to consider are the single-trace half-BPS operators given by
\beq\label{def_halfBPS}
\cO_p(x,y) = y^{i_1}\ldots y^{i_p} {\rm Tr}\left( \phi_{i_1}(x)\ldots \phi_{i_p}(x) \right),\qquad  y \cdot y = 0\ ,
\eeq 
where $y^{i}$ is a complex null vector parametrizing the coset space $SU(4)/ S(U(2)\times U(2))$. For $p=2$ the above operator is the superconformal primary of the energy-momentum multiplet which is dual to the graviton multiplet in the ${\rm AdS}_5$ supergravity theory. For higher $p$ the operator is dual to Kaluza-Klein modes associated to the $S^5$ factor of the ten-dimensional background.

To discuss four-point functions is it helpful to introduce the propagator
\be
g_{ij} = \frac{y_{ij}^2}{x_{ij}^2}\,, \qquad y_{ij}^2 = y_i \cdot y_j\,,
\ee
and conformal cross ratios for both the $x$ and $y$ variables 
\begin{align}
u &= x \bar{x} = \frac{x_{12}^2 x_{34}^2}{x_{13}^2 x_{24}^2}\,, \qquad v = (1-x)(1-\bar{x}) = \frac{x_{14}^2 x_{23}^2}{x_{13}^2 x_{24}^2}\,,\\
\frac{1}{\sigma} &= y \bar{y} = \frac{y_{12}^2 y_{34}^2}{y_{13}^2 y_{24}^2}\,, \qquad \frac{\tau}{\sigma} = (1-y)(1-\bar{y}) = \frac{y_{14}^2 y_{23}^2}{y_{13}^2 y_{24}^2}\,.
\end{align}

Let us now consider the correlators $\langle \mathcal{O}_2 \mathcal{O}_2 \mathcal{O}_3 \mathcal{O}_3 \rangle$ and $\langle \mathcal{O}_2 \mathcal{O}_3 \mathcal{O}_2 \mathcal{O}_3 \rangle$, corresponding to AdS amplitudes of two graviton multiplets and two Kaluza-Klein modes. We write each correlation function as sum of its free theory contribution and an interacting term, 
\begin{align}
\label{freeintdecomp}
\langle \mathcal{O}_2 \mathcal{O}_3 \mathcal{O}_2 \mathcal{O}_3 \rangle 
&=\langle \mathcal{O}_2 \mathcal{O}_3 \mathcal{O}_2 \mathcal{O}_3 \rangle_{\rm free} + \langle \mathcal{O}_2 \mathcal{O}_3 \mathcal{O}_2 \mathcal{O}_3 \rangle_{\rm int}\,,\notag\\[.2cm]
\langle \mathcal{O}_2 \mathcal{O}_2 \mathcal{O}_3 \mathcal{O}_3 \rangle 
&=\langle \mathcal{O}_2 \mathcal{O}_2 \mathcal{O}_3 \mathcal{O}_3 \rangle_{\rm free} + \langle \mathcal{O}_2 \mathcal{O}_2 \mathcal{O}_3 \mathcal{O}_3 \rangle_{\rm int}\,.
\end{align}
Due to the property of partial non-renormalisation \cite{Eden:2000bk} the interacting parts have the following structure 
\begin{align}
\langle \mathcal{O}_2 \mathcal{O}_3 \mathcal{O}_2 \mathcal{O}_3 \rangle_{\rm int} &= g_{12}^2 g_{34}^2 g_{24} \, \mathcal{I}(u,v;\sigma,\tau) \, u^2 G(u,v)\,, \\[.2cm]
\langle \mathcal{O}_2 \mathcal{O}_2 \mathcal{O}_3 \mathcal{O}_3 \rangle_{\rm int} &= g_{12}^2 g_{34}^3 \, \mathcal{I}(u,v;\sigma,\tau)\, u^2 F(u,v)\,, 
\end{align}
where $\mathcal{I}$ is given by
\be
\mathcal{I}(u,v;\sigma,\tau) = \Bigl(\frac{x}{y}-1\Bigr)\Bigl(\frac{x}{\bar{y}}-1\Bigr)\Bigl(\frac{\bar{x}}{y}-1\Bigr)\Bigl(\frac{\bar{x}}{\bar{y}}-1\Bigr) = \frac{s(x,\bar{x};y,\bar{y})}{(y\bar{y})^2}\,.
\ee
The dependence of the correlators on the gauge coupling is entirely through the functions $F(u,v)$ and $G(u,v)$. 

Crossing transformations relate the two correlators and hence the two functions $F(u,v)$,
\be
\label{FGrel}
F(u,v) = \frac{1}{u^4} G\biggl(\frac{1}{u},\frac{v}{u}\biggr)\,. 
\ee
Since we have pairs of identical operators in the correlator we have the symmetry 
\be\label{transf_prop}
G(v,u)=G(u,v)\,, \qquad F\biggl(\frac{u}{v},\frac{1}{v}\biggr) = v^{4} F(u,v)\,.
\ee

The perturbative expansion in string theory or supergravity corresponds to an expansion of the correlators for large $N$. As in \cite{Aprile:2017bgs}, we choose for convenience the expansion parameter
\be
a=\frac{1}{N^2-1}\,.
\ee
With the above choice, the free theory correlation function then has exactly two terms which we express as follows,
\begin{align}
\label{freetheory}
\langle \mathcal{O}_2 \mathcal{O}_2 \mathcal{O}_3 \mathcal{O}_3 \rangle_{\rm free} &= A\Bigl(\langle \mathcal{O}_2 \mathcal{O}_2 \mathcal{O}_3 \mathcal{O}_3 \rangle_{\rm free}^{(0)} + a \langle \mathcal{O}_2 \mathcal{O}_2 \mathcal{O}_3 \mathcal{O}_3 \rangle_{\rm free}^{(1)}\Bigr) \,, \notag \\
\langle \mathcal{O}_2 \mathcal{O}_3 \mathcal{O}_2 \mathcal{O}_3 \rangle_{\rm free} &= A\Bigl(\langle \mathcal{O}_2 \mathcal{O}_3 \mathcal{O}_2 \mathcal{O}_3 \rangle_{\rm free}^{(0)} + a \langle \mathcal{O}_2 \mathcal{O}_3 \mathcal{O}_2 \mathcal{O}_3 \rangle_{\rm free}^{(1)}\Bigr)\,,
\end{align}
where the $N$-dependent factor 
\be
A = \frac{(N^2-1)^2(N^2-4)}{N}
\ee
 has been extracted so that the remaining factor is finite in the large $N$ limit.
Explicitly we have
\begin{align}
\langle \mathcal{O}_2 \mathcal{O}_2 \mathcal{O}_3 \mathcal{O}_3 \rangle_{\rm free} = 6A\Bigl(g_{12}^2 g_{34}^3 + 6 a \bigl( g_{12} g_{34}^2 g_{13} g_{24} +  g_{12} g_{34}^2 g_{14} g_{23} + 2 g_{34} g_{13} g_{24} g_{14} g_{23} \bigr) \Bigr)
\end{align}
with $\langle \mathcal{O}_2 \mathcal{O}_3 \mathcal{O}_2 \mathcal{O}_3 \rangle_{\rm free}$ obtained by crossing.
The interacting parts, or equivalently the functions $F(u,v)$ and $G(u,v)$, have expansions of the form
\begin{align}
\label{aexpansions}
\langle \mathcal{O}_2 \mathcal{O}_2 \mathcal{O}_3 \mathcal{O}_3 \rangle_{\rm int} = A \sum_{n=1}^{\infty} a^n \langle \mathcal{O}_2 \mathcal{O}_2 \mathcal{O}_3 \mathcal{O}_3 \rangle_{\rm int}^{(n)} \,,\qquad F(u,v) = A \sum_{n=1}^{\infty} a^n F^{(n)}(u,v)\,, \notag \\
\langle \mathcal{O}_2 \mathcal{O}_3 \mathcal{O}_2 \mathcal{O}_3 \rangle_{\rm int} = A \sum_{n=1}^{\infty} a^n \langle \mathcal{O}_2 \mathcal{O}_3 \mathcal{O}_2 \mathcal{O}_3 \rangle_{\rm int}^{(n)} \,,\qquad G(u,v) = A \sum_{n=1}^{\infty} a^n G^{(n)}(u,v)\,.
\end{align}
In terms of the string loop expansion the order $a^0$ terms constitute the disconnected contributions to the amplitudes, the order $a$ terms correspond to tree-level connected contributions while order $a^2$ terms correspond to one-loop corrections and so on. In terms of the decomposition (\ref{freeintdecomp}) the order $a$ terms are special, in that they receive contributions from both free theory and from the interacting part of the correlator.

Finally at each perturbative order in $a$ we may expand $F^{(n)}$ and $G^{(n)}$ in powers of $\log u$ multiplied by coefficients analytic at $u=0$,
\be
\label{FGlogexp}
F^{(n)}(u,v) = \sum_{r=0}^n \log^r u \, F^{(n)}_r (u,v)\,, \qquad G^{(n)}(u,v) = \sum_{r=0}^n \log^r u \, G^{(n)}_r (u,v)
\ee
This then makes the branch cut structure around $u=0$ manifest. In particular from OPE considerations we expect that the leading discontinuity at order $a^n$ is of the form $\log^n u $.

\section{Overview of the OPE and double-trace spectrum}
\label{sec_Overview}

Let us consider the contribution of a conformal primary operator $K_{\Delta,l}$ of dimension $\Delta$ and spin $l$ to the OPE of two half BPS operators $\cO_{p_1}$ and $\cO_{p_2}$. It is given as 
\begin{align}
\label{OPE}
	\cO_{p_1}(x_1)\cO_{p_2}(x_2) &\sim  C_{p_1p_2;K_{\Delta,l}}(a)(x_{12}^2)^{\frac{\Delta(a)-l}2-p_1-p_2}  \, \, x_{12}^l K_{\Delta,l}+ \dots\,,
\end{align}
where the dots denote contributions from descendant operators. In the above we have suppressed the representations of the $SU(4)$ global symmetry but we have made explicit the fact that the dimension $\Delta$ and OPE coefficients $C_{p_1p_2;K_{\Delta,l}}$ depend on our expansion parameter, in this case $a$. The quantities $\Delta$ and $C$ therefore admit perturbative expansions,
\begin{align}
\Delta(a) &= \Delta^{(0)} + 2 a \eta^{(1)} + 2 a^2 \eta^{(2)} + \ldots \, \notag \\
C_{p_1p_2;K_{\Delta,l}}(a) &= C_{p_1p_2;K_{\Delta,l}}^{(0)} + a C_{p_1p_2;K_{\Delta,l}}^{(1)} + \ldots\,.
\end{align}
The OPE (\ref{OPE}) is a fully non-perturbative relation, but when expanded perturbatively in $a$ it implies that at $O(a)$ the operator contributes to the discontinuity in $x_{12}^2$ as follows,
	\begin{align}
	\cO_{p_1}(x_1)\cO_{p_2}(x_2) &\ \sim \ a\, C_{p_1p_2;K_{\Delta,l}}^{(0)}\ \eta^{(1)} \, \log x_{12}^2 \   (x_{12}^2)^{\frac{\Delta(0)-l}2-p_1-p_2}  \, \, x_{12}^l K_{\Delta,l}+ \dots.	
	\end{align}
while at $O(a^2)$ it contributes to the double discontinuity as,
	\begin{align}
	\cO_{p_1}(x_1)\cO_{p_2}(x_2) &\ \sim \ a^2\, C_{p_1p_2;K_{\Delta,l}}^{(0)}\ \tfrac{1}{2} (\eta^{(1)})^2\,  \log^2 x_{12}^2\   (x_{12}^2)^{\frac{\Delta(0)-l}2-p_1-p_2}  \, \, x_{12}^l K_{\Delta,l}+ \dots.	
	\end{align}
In the context of four point correlation functions we see that at order $a^2$ the double discontinuity in $x_{12}^2$ (and hence in the conformal cross-ratio $u$) comes entirely from zeroth order OPE coefficients and first order anomalous dimensions $\eta^{(1)}$ of $K_{\Delta,l}$. These same quantities are already present in the single discontinuity at order $a$.

In the supergravity regime of $\mathcal{N}=4$ SYM, we wish to bootstrap the correlator $\langle \cO_2\cO_2\cO_3\cO_3 \rangle$ 
at one-loop level, from known lower order results. To achieve this we use the fact described above that the double discontinuity in $u$ at order $a^2$ 
depends entirely on the zeroth order OPE coefficients and first order anomalous dimensions.  We take into account the contributions of all superconformal descendants by making use of the superconformal partial wave (SCPW) expansion of the correlation function.
Doing so we find the double discontinuity of the correlator at order $a^2$ is given by
\begin{align}\label{eq1_log2u}
		\langle \cO_2\cO_2\cO_3\cO_3 \rangle^{(2)}\Big|_{\log^2 u} &= g_{12}^2 g_{34}^3\, \mathcal{I}(u,v;\sigma,\tau) \,u^2 F^{(2)}_2(u,v) \notag \\ 
		&=  g_{12}^2 g_{34}^3\, \frac{1}{2} \sum_{t,l,i}\, \bigl(\eta^{(1)}_{t,l,i}\bigr)^2 \bigl( C^{(0)}_{22;K_{t,l,i}} C^{(0)}_{33;K_{t,l,i}}\bigr) \, \LL^{2233}_{[0,0,0]}(t,l)\,. 
\end{align}
Here $\LL_{[0,0,0]}(t,l)$ are long superconformal blocks (for the precise definition see eq. (\ref{longmultiplet})) corresponding to the exchange of long double trace multiplets with $SU(4)$-singlet superconformal primary operators $K_{t,l,i}$,
\begin{align}\label{op_schann}
		\{K_{t,l,1},K_{t,l,2},\dots K_{t,l,t{-}1}\} &\sim \{ \cO_2 \partial^l \Box^{t-2} \cO_2,\cO_3 \partial^l \Box^{t-3} \cO_3, \dots ,\cO_t \partial^l \cO_t \}|_{[0,0,0]}\,.
\end{align}
These operators are degenerate in the large $N$ limit with dimension $\Delta^{(0)} = 2t+l$
but acquire non trivial anomalous dimensions $\eta^{(1)}_{t,l,i}$ at subleading order in $a$.

In principle more operators could contribute to the OPE, 
but in the supergravity limit the space of operators is significantly simplified\footnote{A key assumption we make is that triple (or higher) trace operators do not contribute to the leading order (in $a$) OPE coefficients}. The double discontinuity in $x_{12}^2$ then 
comes entirely from zeroth order three-point functions and the anomalous dimensions. In order to determine this data, 
we have to take into account an important subtlety: we can not determine $C^{(0)}_{22;K_{t,l,i}}$, $C^{(0)}_{33;K_{t,l,i}}$, and $\eta_{t,l,i}^{(1)}$ individually 
from a superconformal partial wave expansion of $\langle \cO_2\cO_2\cO_3\cO_3 \rangle^{(0)}$ and $\langle \cO_2\cO_2\cO_3\cO_3 \rangle^{(1)}$,  but only
\be
\sum_i C^{(0)}_{22;K_{t,l,i}} C^{(0)}_{33;K_{t,l,i}}
\ee
and 
\be
\sum_i \eta_{t,l,i}^{(1)} \,C^{(0)}_{22;K_{t,l,i}} C^{(0)}_{33;K_{t,l,i}}\,.
\ee 
However, this problem can be overcome and in fact it has been explicitly solved in~\cite{unmixing} by considering
the more general family of correlators $\langle \cO_p\cO_p\cO_q\cO_q \rangle$. 
For more details, we refer the reader to that paper. We will recall the formulas obtained from that analysis in Sect. \ref{sect-anomdimsOPEcoeffs} as we will need them to explicitly construct the double discontinuity of $\langle \mathcal{O}_2 \mathcal{O}_3 \mathcal{O}_2 \mathcal{O}_3 \rangle$.

In the case of $\langle \cO_2\cO_2\cO_3\cO_3 \rangle$, the double discontinuity in the channel $x_{12}^2\rightarrow 0$ 
is not enough to attempt to determine the full correlator. For illustration, let us consider the one loop result for the correlator 
$\langle \cO_2\cO_2\cO_2\cO_2 \rangle$, which has been obtained in \cite{Aprile:2017bgs}.
In that case, we found that the part of the correlator of transcendental weight four is determined by the sum of the double box function 
in three different orientations. One of these orientations contains no double discontinuity in the limit $x_{12}\rightarrow 0$. 
When the external operators have equal charges, like $\langle \cO_2\cO_2\cO_2\cO_2 \rangle$, 
crossing symmetry relates the three orientations, but for $\langle\cO_2\cO_2\cO_3\cO_3 \rangle$ we would never detect its coefficient.
Therefore, we need to consider double discontinuities in all possible channels, in particular we need to consider 
an inequivalent OPE limit $x_{13}^2\rightarrow 0$. This is the same as considering the correlator 
$\langle \cO_2\cO_3\cO_2\cO_3 \rangle$ in the limit $x_{12}^2\rightarrow 0$.  The study of the OPE in this channel 
will be slightly more involved compared to $\langle \cO_2\cO_2\cO_3\cO_3 \rangle$. The long double trace operators 
which we need to consider are given by
\begin{align}\label{op_tchann}
\{\cK_{t,l,1},\cK_{t,l,2},\dots \cK_{t,l,t{-}1}\} &\sim \{ \cO_2 \partial^l \Box^{t-2} \cO_3,\cO_3 \partial^l \Box^{t-3} \cO_{4}, \dots ,\cO_t \partial^l \cO_{t+1} \}|_{[0,1,0]}
\end{align}
where the basis of operators on the l.h.s. is characterized by having odd twist and both even and odd spins. 
Expanding their dimensions and three point function coefficients $\langle \cO_2\cO_3\cK_{t,l,i}\rangle$ as follows, 
\bea
\Delta_{t,l,j}&=& 2t+1 + l +2a \eta^{(1)}_{t,l,j} + 2a^2 \eta_{t,l,j}^{(2)}+\ldots\\
C_{23;\cK_{t,l,j}}&=&
					 C^{(0)}_{23;\cK_{t,l,j}} + a C^{(1)}_{23;\cK_{t,l,j}}+\ldots 
\eea
we readily obtain the result for double discontinuity at order $1/N^4$, i.e. 
\begin{align}\label{double_disc_2323}
\langle \cO_2\cO_3\cO_2\cO_3 \rangle^{(2)}\Big|_{\log^2 u} &= g_{12}^2 g_{34}^2 g_{24} \, \mathcal{I}(u,v;\sigma,\tau) \, u^2 G^{(2)}_2(u,v) \notag \\  
&=g^2_{12}g^2_{34}g_{24}\, \frac{1}{2} \sum_{t,l,j}\, \bigl(\eta^{(1)}_{t,l,j}\bigr)^2 \bigl(C^{(0)}_{23;\cK_{t,l,j}}\bigr)^2  \LL^{2323}_{[0,1,0]}(t+\tfrac{1}{2} |l) 
\end{align}
where the long superblocks now correspond to the operators in \eqref{op_tchann}. These operators are again degenerate in the large $N$ limit, thus in order to bootstrap the double discontinuity in \eqref{double_disc_2323}, we have to solve a new mixing problem.


\section{Unmixing in $[0,1,0]$} 

\label{sec4}

The problem of operator mixing in the $[0,1,0]$ representation is similar but not completely analogous 
to that of the $[0,0,0]$ representation. The correlators we have to 
consider are of the form $\langle \cO_p \cO_{p+1} \cO_q \cO_{q+1}\rangle$. The relevant terms in the SCPW expansion, 
restricted to the exchange of long multiplets, are
\begin{align}
\langle \cO_p \cO_{p+1} \cO_q \cO_{q+1}\rangle_{\rm long}=  N^{ \Sigma } \, g_{12}^p g_{34}^q g_{24} &  \nonumber
									\Bigg( \sum_{t,\,l,\, \mathfrak{R}} \cA^{\{p,q\}}_{\,\mathfrak{R}}(t|l)\  \LL^{\{p,q\}}_{\, \mathfrak{R} }(t+\tfrac{1}{2}|l) \label{unmixing1}\\[.1cm]
 &\rule{.3cm}{0pt}
+\frac{1}{N^2}\log(u)\, \sum_{t,\, l,\,\mathfrak{R}} \symbol^{\{p,q\}}_{\,\mathfrak{R}}(t|l)\ \LL^{\{p,q\}}_{\, \mathfrak{R} }(t+\tfrac{1}{2}|l)\, +\,  \ldots \rule{.2cm}{0pt} \Bigg) 
\end{align}
where $\Sigma=p+q+1$ and the dots refer to terms analytic at $u=0$ as well as terms of higher order in $1/N^2$. The coefficients $\cA^{\{p,q\}}_{\,\mathfrak{R}}(t|l)$ and $\symbol^{\{p,q\}}_{\,\mathfrak{R}}(t|l)$ 
are obtained from disconnected free theory and tree level supergravity, respectively. The corresponding long superblocks 
$\LL^{\{p,q\}}$ will be given explicitly in the next section. From the OPE and the knowledge of the spectrum 
of double trace operators $\cK_{t,l,i}$ described in (\ref{op_tchann}), we deduce the two equations, 
\begin{eqnarray}
	 	  \cA^{\{p,q\}}_{[0,1,0]}(t|l)		&=&		\sum_{i=1}^{t-1} C^{(0)}_{p\,p+1;\cK_{t,l,i}} \,C^{(0)}_{q\,q+1;\cK_{t,l,i}}		  \label{mixing_equation1} \\
		  \symbol^{\{p,q\}}_{[0,1,0]}(t|l)	&=&		\sum_{i=1}^{t-1} \eta_i^{(1)}\, C^{(0)}_{p\,p+1;\cK_{t,l,i}}  C^{(0)}_{q\,q+1;\cK_{t,l,i}}  \ .\label{mixing_equation2}
\end{eqnarray} 
In the following we will drop the superscript $(0)$ and $(1)$ since there is no ambiguity at this order. 
We now prove that the set of OPE coefficients $C_{p\,p+1;\cK_{t,l,i}}$ 
and anomalous dimensions $\eta_i$  is uniquely specified by the solution of these two equations.
In fact, for given twist and spin, $C_{p\,p+1;\cK_{t,l,i}}$ is non-zero only when $2\leq p\leq t$, and by taking into 
account the $p \leftrightarrow q$ symmetry, we conclude that the l.h.s of \eqref{mixing_equation1}-\eqref{mixing_equation2} 
determines $t(t-1)$ independent pieces of data. The number of unknowns, on the other hand, is given by $t-1$ 
anomalous dimensions  $\eta_i$ together with $(t-1)^2$ OPE coefficients $C_{p\,p+1;\cK_{t,l,i}}$ (because $i$ runs from 1 to $t-1$ and $p$ from 2 to $t$). 
Thus there are a total of $t(t-1)$ unknowns, exactly the same as the number of independent CPW coefficients. 

As we mentioned, the SCPW expansion \eqref{unmixing1} contains both even and odd spins, and furthermore the sum over twist runs over odd integers. 
Compared to the study of $\langle \cO_p \cO_{p} \cO_q \cO_{q}\rangle$ we then expect some differences, and we will show that
the unmixing is modified in an interesting way.


\subsection{Disconnected Free Theory}

In the first instance we are interested in the leading large $N$ contribution in the correlators $\langle \cO_p\cO_{p+1}\cO_q\cO_{q+1}\rangle$. 
The leading large $N$ contribution comes from the disconnected diagrams, i.e. the contribution to the four-point function which factorises into a product of two-point functions. Since the operators are protected, the two-point functions are independent of the 't Hooft coupling and take their large $N$ free-field forms.

We can therefore consider the various free-field propagator structures in $\langle \cO_p\cO_{p+1}\cO_q\cO_{q+1}\rangle$ and isolate the disconnected one. Such terms are only present for $p=q$ while the case $p\neq q$ is subleading at large $N$, 
\beq
\langle \cO_p \cO_{p+1} \cO_q \cO_{q+1} \rangle =  N^{2p+1} \left( \delta_{pq} p (p+1) g_{13}^p g_{24}^{p+1} + O(1/N^2)\right)\ .
\label{corpp1qqp1}
\eeq
The general expression for the superconformal partial wave expansion at leading order for large $N$ is then given by
\begin{align}
\langle \cO_p \cO_{p+1} \cO_p \cO_{p+1} \rangle&=\mathcal{P}^{\rm\, OPE} \sum_{\gamma,\, \ula}  A_{\gamma,\ula} \,  \mathbb{S}^{\alpha\beta\gamma;\ula}\,,\label{corblock}
\end{align}
where $\mathcal{P}^{\rm\, OPE}= g_{12}^p g_{34}^p g_{24}$ and $\mathbb{S}^{\alpha\beta\gamma;\ula}$ are superconformal blocks \cite{Dolan:2000ut,Dolan:2001tt,Dolan:2004iy,Dolan:2006ec,Doobary:2015gia}.
We follow the notation introduced in \cite{Doobary:2015gia}, where the superblocks are specified by three integers $\alpha,\beta,\gamma$ and a Young tableau $\ula$. For the specific case under consideration, $\alpha=\tfrac{\gamma+1}{2}$, $\beta=\tfrac{\gamma-1}{2}$, and
the superconfomal block is given by the following determinantal formula 
\begin{align}\label{detform}
\mathbb{S}^{\alpha\beta\gamma;\ula} &= 
								\left(\frac{x \bar{x}}{y \bar{y}}\right)^{\frac12(\gamma-1)} 
								F^{\alpha\beta\gamma\ula}\qquad 
								\gamma = 1,3, \dots, 2p+1 \notag\\[.2cm]
F^{\alpha\beta\gamma\ula}\ &=\  (-1)^{ p + 1} \frac{s(x,\bar{x},y,\bar{y})}{(x-\bar{x})(y-\bar{y})}
\det \left( \begin{array}{cc}
						F^X_{\underline\lambda}&	   R		\\
						K_{\underline \lambda}   &   F^Y
		\end{array}\right)\ , \notag\\[.2cm]
s(x,\bar{x},y,\bar{y}) &= { (x-y)(x-\bar{y})(\bar{x}-y)(\bar{x}-\bar{y})}\,, 
\end{align}
Precise definition of the determinantal formula can be found in Appendix~\ref{Appendix--SCPW}. 
Here we are interested in the coeffiecients $A_{2p+1,\ula}$ corresponding to long multiplets with twist $2t+1$, spin $l$ and  $SU(4)$ representation $[0,1,0]$.

This translates to a Young tableau with row lengths $\ula=[t{-}p{+}l{+}2,t{-}p{+}2,2^{p-2}]$, as can be read off using the table in the appendix, eq~\eqref{tableapp}. 
Inputting the correlator~\eqref{corpp1qqp1} and the superblocks~\eqref{detform}, with the relevant value $\gamma=2p+1$, the equation~\eqref{corblock} reduces to
\begin{align}
	N^{2p+1} p (1{+}p) = \sum_{\ula}  A_{2p+1,\ula}   F^{p,p+1,2p+1;\ula}\ .
\end{align}
Here, the left hand side is a constant, where as the right hand side is a function of $x,\bar x,y,\bar y$. There is a unique solution, yielding the values of the coefficients $A_{2p+1,\ula}$.

In fact, there is a conceptually simpler way to solve this equation. As outlined in~\cite{unmixing} the entire superblock formalism can be bosonised. Since the Young tableau has height $p$ we can use bosonised $GL(p,p)$ blocks described in~\cite{Doobary:2015gia,unmixing} for which%
\footnote{In this formula (only) we use $x_i$ where $i=1\dots p$ to represent generalised cross-ratios in the $GL(p,p)$ theory. These should be thought of as generalisations of the two independent cross-ratios $x,\bar x$ of a conformal theory in 4d, corresponding to $p=2$. They should not be confused with space-time coordinates.}
  \begin{align}\label{eq:25}
  F^{\alpha\beta\gamma\underline\lambda}(\underline x) &=\frac{\det\Big(
  	x_i^{\lambda_j+p-j}{}_2F_1(\lambda_j{+}1{-}j{+}\alpha ,\lambda_j{+}1{-}j{+}\beta;2\lambda_j{+}2{-}2j{+}\gamma;x_i)\Big)_{1\leq i,j \leq
  		p}}{\det\Big(
  	x_i^{p-j}\Big)_{1\leq i,j \leq
  		p}}\ .
  \end{align}
  The advantage here, is that on does not have to deal with the different cases needed for short superblocks, but can use one formula to deal with all the blocks.
  
With either method (superblocks or bosonised blocks) the resulting block coefficients   are consistent with the formula
\bea\label{App+1pp+1}
\cA^{ \{ p\,p{+}1\,p\,p{+}1\} }_{[0,1,0]} &=\frac{ 72 ((t+1)!)^2  (t+4)_{p-2} (t-p+1)_{p-2}  ((l+t+2)!)^2   (l+1) (l+2 t+3)(l+t+5)_{p-2} (l-p+t+2)_{p-2}}{ (p-2)!((p-1)!)^2 (p+2)!  (2 t+1)! (2 l+2 t+3)!}\ .
\eea
For given twist $2t+1$ and spin $\ell$, we can finally assemble the data into the diagonal matrix
\beq
\widehat{\cA}(t|l)={\rm diag}\left( \cA_{[0,1,0]}^{\{ 2,3,2,3\} },\ldots ,\cA^{\{t,t+1,t,t+1\}}_{[0,1,0]}\right)\ .
\eeq


\subsection{Tree Level Supergravity from Mellin space}

The supergravity amplitude for correlators of the form $\langle \cO_p \cO_{p+1} \cO_q \cO_{q+1}\rangle$ can be obtained from 
the general result of Rastelli and Zhou \cite{Rastelli:2016nze}. Their formula extends in a consistent way the very few 
explicit computations of Witten diagrams known in the literature \cite{hep-th/9903196,Arutyunov:2000py,Uruchurtu:2008kp,Uruchurtu:2011wh}, and gives access to correlators with arbitrary configuration of charges.
These are precisely the correlators we need in order to obtain $\symbol^{\{p,p+1,q,q+1\}}$ and solve the mixing problem. 
As an example, in Appendix~\ref{Appendix--ExamplesSugra} we compute  
the Mellin amplitude corresponding to $q=2,3,4,5$ with $p\ge q$ generic, and we rewrite it in a standard basis of $\Db_{\delta_1\delta_2\delta_3\delta_4}$ functions. The simplest case is,\footnote{
We recall the identity  $$s(x,\bar{x};y,\bar{y})/(y\bar{y})^2= v+ \sigma^2 u v + \tau^2 u + \sigma v (v-1-u) + \tau (1-u-v)+\sigma\tau u (u-1-v).$$}
\bea 
\langle \cO_p \cO_{p+1} \cO_2 \cO_{3}\rangle_\text{sugra}&=& g_{12}^p g_{34}^2 g_{24} \frac{ s(x,\bar{x};y,\bar{y})}{(y\bar{y})^2}\ \norma_{p,p+1,2,3}\,\cH^{\rm int}_{p,p+1,2,3} \\[.2cm]
\cH_{p,p+1,2,3}^{\rm int}&=& \frac{u^p}{v}\Db_{p+3,p,3,2}
\eea
In general, $\cH_{p,p+1,q,q+1}$ is a polynomial of degree $p-2$ in the $SU(4)$ variables.

At order $1/N^2$ the full  $\langle \cO_p \cO_{p+1} \cO_q \cO_{q+1}\rangle$ correlator contains two contributions: 
supergravity and connected free theory. The knowledge of the Mellin amplitude does not fix the supergravity correlator completely, 
and we will have to determine the relative normalisation $\norma_{p,p+1,q,q+1}$ by an independent argument.  
For correlators of the form $\langle \cO_p \cO_{p} \cO_q \cO_{q}\rangle$ we obtained the corresponding normalisation by considering the absence 
of twist 2 long operators in the spectrum of supergravity \cite{unmixing}. For equal charges, these results can be used to predict the normalisation 
of $\langle \cO_p \cO_{p+1} \cO_p \cO_{p+1}\rangle$ and show that twist 3 long conformal partial waves cancel between free theory and supergravity. 
Then, we can obtain the normalisation $\norma_{p,p+1,q,q+1}$ by imposing the absence of twist 3 long operators at order $1/N^2$.

The absence of twist 3 long operators is not immediately transparent, since the twist 3 short (half-BPS) operator remains, and so its conformal block is present. A very simple way to avoid this technicality is to project onto the large spin limit of the twist 3 operators. The twist 3 operators correspond to the $u\rightarrow 0$ limit of the correlator. Further taking $v\rightarrow 0$ then projects onto the large spin limit. The advantage of this is that then one doesn't have to deal with the twist 3 short operator but can simply insist on the vanishing of the two contributions in the limit $u,v \rightarrow 0$. Note that this limit corresponds to taking the light-like polygonal limit relevant for the duality with Wilson loops and amplitudes in $\mathcal{N}=4$ SYM.

The free theory propagator structure at subleading order in $N$ contains 
\be
 \langle \cO_p\, \cO_{p+1}\, \cO_{q}\,\cO_{q+1} \rangle^{(1)}_{\rm free}=   N^{\Sigma-2}\Big( 2 p(p+1)q(q+1)  g_{12}^{p-1} g_{34}^{q-1} g_{24}g_{14}g_{23}  + \ldots \Big)
\ee
where  $\Sigma=p+q+1$, and we omitted propagators structures contributing to higher twist CPW, i.e not leading in the $u\rightarrow 0$ expansion.\footnote{
	Note that one might expect a term  proportional to $g_{12}^p g_{34}^q g_{24}$ which would contribute to lower twist, but its coefficient vanishes due to properties of $SU(N)$ vertices. } The limit $v\rightarrow 0$ with $u/v$ fixed gives 
\begin{align}\label{limit_twist_3}
\lim_{u,v\rightarrow 0} 
\frac{ \langle \cO_p\, \cO_{p+1}\, \cO_{q}\,\cO_{q+1} \rangle^{(1)} }{  g_{12}^{p} g_{34}^{q} g_{24}}\Big|_\text{free}\  =  \ \frac{2 p(p+1)q(q+1)}{N^2}  \frac{g_{14}g_{23}}{g_{12}g_{34}}= \frac{2 p(p+1)q(q+1)}{N^2} \frac{\tau u}{v}\, .
\end{align}
Notice that we factorized $\cP^{\rm OPE}$ in the denominator on the l.h.s. 
Given the generic form of the supergravity correlator with $p\ge q$, 
\bea \label{limit_twist_3_sugra}
\langle \cO_p \cO_{p+1} \cO_q \cO_{q+1}\rangle_\text{sugra}&=& g_{12}^p g_{34}^q g_{24} \frac{ s(x,\bar{x};y,\bar{y})}{(y\bar{y})^2}\ \norma_{p,p+1,q,q+1}\,\cH^{\rm int}_{p,p+1,q,q+1} 
\eea
we would like to take the limit $u,v\rightarrow 0$ as we did in the corresponding free theory. 
The dynamical function $\cH^{\rm int}_{p,p+1,q,q+1}$ has in general non trivial dependence on the $SU(4)$ variables, however we find that 
$
\lim_{u,v\rightarrow 0}  s(x,\bar{x},y,\bar{y})/(y\bar{y})^2 =\tau\,,
$
therefore in order to match the r.h.s of eq.~\eqref{limit_twist_3}, it is sufficient to consider the leading term in $u/v$ of $\mathcal{H}_{p,p+1,q,q+1}$ restricted to $\sigma=\tau=0$. 
It can be inferred from the expression of its Mellin amplitude, and explicitly checked in the examples \eqref{corr_23}-\eqref{corr_45} and \eqref{corr_56}, that 
\beq
\mathcal{H}_{p,p+1,q,q+1}\Big|_{\sigma=\tau=0}= \frac{(q-2)! (p-q)! }{(p-2)!} \sum_{k=0}^{p-2} \frac{1}{k!} \frac{1}{v}\, u^p\, \Db_{p+3,p,3+k,2+k}(u,v),
\eeq
The representation of $\Db_{p_1,p_2,p_3,p_4}$ contains three different analytic contributions,
\beq\label{DbaruY2}
\Db_{\delta_1\delta_2\delta_3\delta_4}= 
			u^{-\sigma}\, \Db_{\delta_1\delta_2\delta_3\delta_4}^{\rm\, sing} +  
						\Db_{\delta_1\delta_2\delta_3\delta_4}^{\rm\, analytic}+\,   
				  \log(u)\,\Db_{\delta_1\delta_2\delta_3\delta_4}^{\rm\, log}\ , 
\eeq
and it is useful to consider them separately. The precise form of these functions is given in Appendix~\ref{Appendix--AnatomyOfDbarfunctions}. 
Here it is enough to recall that for the relevant values of $\delta_{i=1,2,3,4}$ the functions $\Db_{\delta_1\delta_2\delta_3\delta_4}^{\rm\, analytic}$, 
$\Db_{\delta_1\delta_2\delta_3\delta_4}^{\rm\, log}$ and $\Db_{\delta_1\delta_2\delta_3\delta_4}^{\rm\, sing}$ 
are analytic in $u$ and therefore 
\beq
\lim_{u\rightarrow 0} u^p \Db_{\delta_1\delta_2\delta_3\delta_4}^{\rm\, analytic}= \lim_{u\rightarrow 0}  u^p\log u\, \Db_{\delta_1\delta_2\delta_3\delta_4}^{\rm\, log}=0.
\eeq
Since $\sigma=(\delta_1+\delta_2-\delta_3-\delta_4)/2$, 
the limit of $u^{-\sigma}\Db^{\rm sing}_{p+3,p,3+k,2+k}$ is more interesting. In our specific case, $\sigma=p-k-1$, and when $k=0$ we obtain the following non trivial result,\footnote{
We repeat for quick reference the expression given in Appendix~\ref{Appendix--AnatomyOfDbarfunctions} \eqref{Dsingularseries}
$$
\Dbar_{\delta_1\delta_2\delta_3\delta_4}^{\rm\,sing}= 
									\sum_{n=0}^{\sigma-1} \frac{(-u)^{n}}{n!}\ 
									\Gamma[\sigma-n]\ \Lambda^{\delta_3\delta_4}_{\delta_1-\sigma\delta_2-\sigma}(n)\ {F}^{\,\delta_2-\sigma+n | \delta_3+n}_{\, \delta_3+\delta_4+2n}(1-v)\ ,
$$
where  
we defined, 
$ F^{a|b}_c(x)\equiv\,_2 F_1[a,b;c](x)$, and
$\Lambda^{\delta_1\delta_2}_{\delta_3\delta_4}(n)\equiv \Gamma[\delta_1+n]\Gamma[\delta_2+n]\Gamma[\delta_3+n]\Gamma[\delta_4+n] / \Gamma[\delta_1+\delta_2+2n]$.
} 
\beq\label{proof_norma}
\frac{u^p}{v} \Db^{\rm\, sing}_{p+3,p,3,2}= \frac{u}{v}\frac{ \Gamma[p-1]}{2} \,_2F_1[1,3,5;1]+O(u). 
\eeq 
By requiring twist $3$ long cancellation at large spin, we obtain from \eqref{limit_twist_3} and \eqref{limit_twist_3_sugra}, the relation
\be
 \frac{2 p(p+1)q(q+1)}{N^2} +  \frac{2(q-2)! (p-q)! }{(p-2)!}  \Gamma[p-1] \norma_{p,p+1,q,q+1} = 0
\ee
which fixes the value of the normalisation to
\begin{align}
\norma_{ p\, p{+}1\, q\,q{+}1} = - \frac{p(p+1)q(q+1)}{(q-2)!(p-q)!}N^{p+q-1}\ .
\end{align}	
We can now proceed, and compute the superconformal partial wave expansion of the long sector of $\langle \cO_p\, \cO_{p+1}\, \cO_{q}\,\cO_{q+1} \rangle_{sugra}$.

The factorized form of the supergravity correlator implies that only long multiplets contribute. 
Thus, the corresponding Young tableau have two or more rows and two or more columns. 
For an expansion in purely long operators there is no great advantage in using bosonised blocks 
and we will use directly the determinantal formula for the superblocks given in Appendix~\ref{Appendix--SCPW}.
As mentioned before, if $p\ge q$ the function $\cH_{q,q+1,p,p+1}$ is a polynomial of degree $p-2$ in the $SU(4)$ variables. 
Therefore, when $p>2$ we will project onto the $[0,1,0]$ representation. 
Results for $[n,1,n]$ will be presented elsewhere.  For given twist $2t+1$ and spin $\ell$ we construct the matrix
\beq
\widehat{\symbol}(t|l)\Big|_{[0,1,0]}=\left(\begin{array}{cccl} 
						\symbol^{\{2,3,2,3\}} & \symbol^{\{3,4,2,3\}} & \ldots &  \symbol^{\{t,t+1,2,3\}} \\[.2cm]
									       &  \symbol^{\{3,4,3,4\}} & \ldots & \symbol^{\{t,t+1,3,4\}} \\[.3cm]
									       &				  & \ldots & \symbol^{\{t,t+1,t,t+1\}}
		    \end{array}\right),				
\eeq
where we have just given the independent entries in the upper triangular part explicitly.
The coefficients, $\symbol^{\{p,p+1,q,q+1\}}$ have different behaviour for even and odd spins. 
For example, the $t$ and $l$ dependence in the first three cases is, 
\begin{align}
\symbol^{\{2,3,2,3\}}= \left\{\begin{array}{ll}
		-\frac{ 6 (t-1)_2  (t+2)_2  ((t+1)!)^2 (l+t+1)! (l+t+2)! (l+2 t+3) }{ (2 t+1)! (2 l+2 t+3)!}&l \text{ even}\\[.2cm]
		-\frac{ 6 (t-1)_2 (t+2)_2  ((t+1)!)^2 (l+t+1)! (l+t+2)! (l+1)        }{ (2 t+1)! (2 l+2 t+3)!}&l \text{ odd}\end{array}\right. \label{M2323}  \\[.4cm]
\symbol^{\{3,4,2,3\}}=\left\{\begin{array}{ll}
		-\frac{ 3 (t-2)_3 (t+2)_3 ((t+1)!)^2 (l+t+1)!(l+t+2)! (l+2t+3)}{ (2 t+1)! (2 l+2 t+3)!} &l \text{ even}\\[.2cm]
		-\frac{ 3 (t-2)_3 (t+2)_3  ((t+1)!)^2 (l+t+1)! (l+t+2)! (l+1)   }{ (2 t+1)! (2 l+2 t+3)!}&l \text{ odd}\end{array}\right. \label{M3423} 	
\end{align}
and 
\begin{align}
\symbol^{\{3,4,3,4\}}=\left\{\begin{array}{ll}
		-\frac{ 3 (t-2)_3 (t+2)_3 ((t+1)!)^2 (l+t+1)!(l+t+2)! (l+2t+3) (2l^2 + l(3t+7) + 10t^2 +23 t-75) }{ 20 (2 t+1)! (2 l+2 t+3)!} &l \text{ even}\\[.2cm]
		-\frac{ 3 (t-2)_3 (t+2)_3  ((t+1)!)^2 (l+t+1)! (l+t+2)! (l+1)( 2l^2 +l(5t+9)+12t^2+29t-71) }{  20 (2 t+1)! (2 l+2 t+3)!}&l \text{ odd}\end{array}\right. \label{M3434}
\end{align}
Therefore we will study $\widehat{\symbol}(t|l)$ for even and odd spins separately.


\subsection{Anomalous dimensions and OPE coefficients}
\label{sect-anomdimsOPEcoeffs}

Once the matrices $\widehat{\cA}(t|l)$ and $\widehat{\symbol}(t|l)$ have been found,  we can solve for the OPE coefficients and the anomalous dimensions.
As in \cite{unmixing}, it is convenient to introduce the matrix of normalized three-point coefficients, 
\beq
\widetilde{c}(t|l)\equiv  \widehat{\cA}^{-\frac{1}{2}}
 								\left(\begin{array}{lccc}
											C_{23 K_{t,l,1}} & C_{23 K_{t,l,2}}  & \ldots  & C_{23 K_{t,l,t-1} }\\
											C_{34 K_{t,l,1}} & C_{34 K_{t,l,2}} & \ldots & \\
											\ldots & & &\\
											C_{t, t+1, K_{t,l,1}} & & & 
\end{array}\right)
\eeq 
and rewrite equations \eqref{mixing_equation1} and \eqref{mixing_equation2} in matrix form. 
The first set of equations becomes the orthonormality condition $\tilde{c}\, \tilde{c}^{\,T}=\text{Id}_{t-1}$, the second one reduces to the eigenvalue problem, 
\beq\label{rastellimatrix}
\tilde{c}\cdot \text{diag}\left(\eta_1,\ldots,\eta_{t-1}\right)\cdot \tilde{c}\,^T =   \widehat{\cA}^{-\frac{1}{2} }\cdot \widehat{\symbol} \cdot \widehat{\cA}^{-\frac{1}{2} }
\eeq
Then, anomalous dimensions are eigenvalues and the corresponding eigenvectors columns of $\widetilde{c}(t|l)$.
We look at the first few cases explicitly.

\subsubsection{Twist 5}

This case is straightforward as there is only one long operator for each spin, and only one correlator to be considered, namely $\langle 2323 \rangle$. 
The result for the anomalous dimensions is
	\begin{align}
		\eta_1=\left\{\begin{array}{ll}
		-\frac{80}{(1+l)(4+l)}& l \text{ even}\\[.3cm]
		-\frac{80}{(4+l)(7+l)}& l \text{ odd}
		\end{array}\right.
	\end{align}
and $\widetilde{c}(2|l)$ is trivial. The only three-point function coefficient is given by plugging $p=t=2$ into equation~\eqref{App+1pp+1}, giving
	\begin{align}
		C_{23 K_{2,l,1}}^2=\frac{9 (l+1) (l+7) ((l+4)!)^2}{10 (2 l+7)!}\ .
	\end{align}

\subsubsection{Twist 7}

In this case we have a two-dimensional space of long operators. The relevant correlators are $\langle 2323\rangle$, $\langle 2334\rangle$, $\langle 3434\rangle$.
Considering even spins, we find 
\beq
\begin{array}{l}
\eta^{\rm even}_{i=1,2}=\left\{ -\frac{360(l+7)}{(l+1)(l+2)(l+5)} ,-\frac{360}{(l+5)(l+8)} \right\},
\end{array}
~~~\widetilde{c}\,(3|l\ {\rm even })=	\left(\begin{array}{cc}
										\sqrt{\frac{7(l+2)}{6(2l+9)}}  & \sqrt{\frac{5(l+8)}{6(2l+9)}}  \\[.2cm]
										-\sqrt{\frac{5(l+8)}{6(2l+9)}}  & \sqrt{\frac{7(l+2)}{6(2l+9)}}  \\
							\end{array}\right),							
\eeq
whereas for odd spins we obtain
\beq
\begin{array}{l}
\eta^{\rm odd}_{i=1,2}=\left\{ -\frac{360}{(l+2)(l+5)},-\frac{360(l+3)}{(l+5)(l+8)(l+9)} \right\},
\end{array}
~~~\widetilde{c}\,(3|l\ {\rm odd })=	\left(\begin{array}{cc}
										\sqrt{\frac{5(l+2)}{6(2l+11)}}  & \sqrt{\frac{7(l+8)}{6(2l+11)}}  \\[.2cm]
										-\sqrt{\frac{7(l+8)}{6(2l+11)}}  & \sqrt{\frac{5(l+2)}{6(2l+11)}}  \\
							\end{array}\right).
\eeq	
It is interesting to consider how the transformation $l\rightarrow -l-10$ acts on the anomalous dimensions. 
Given the set $\{ \eta_{1}^{\rm even},\eta_{2}^{\rm even},\eta_{1}^{\rm odd},\eta_{2}^{\rm odd}\}$ 
the transformation exchanges $\eta_1^{\rm even}\leftrightarrow\eta_2^{\rm odd}$ and $\eta_2^{\rm even}\leftrightarrow\eta_1^{\rm odd}$. 
As a consequence, the square root structure in the columns of $\widetilde{c}(l\ {\rm even})$ is related to that 
of the columns of $\widetilde{c}(l\ {\rm odd})$ in the same way. (The signs $\pm1$ are fixed by orthogonality independently of the symmetry). Compared to the singlet channel \cite{unmixing}, 
where a similar transformation acted separately within each spin sector, even and odd, the exchange property here is novel, 
and the generalization of $\widetilde{c}(t|\ell)$ is actually less trivial than what we could have naively guessed. We discuss this symmetry in more detail in section~\ref{sec7}.

\subsubsection{Twists 9 and 11}

Before presenting general results for the anomalous dimensions $\eta_i$ and $\widetilde{c}(t|l)$, we give two more examples, twist 9 and 11.
In the first case, the space of long operators is three dimensional. For even spins we have
\beq
\begin{array}{ccl}
&&
\eta^{\rm even}_i = \left\{ -\frac{1008 (l+7) (l+8)}{(l+1) (l+2) (l+3) (l+6)}, -\frac{1008}{(l+3) (l+6)}, -\frac{1008 (l+4)}{(l+6) (l+9) (l+10)}
				     \right\},\\ 
				     \\
				      \\
\widetilde{c}(4|l\ {\rm even})&=&
\left(
\begin{array}{ccc}
\sqrt{  \frac{ 3  (l+2)(l+3) }{ 2 (2 l+9) (2 l+11) } }  &    \sqrt{ \frac{ 5 (l+3)(l+9)}{ 3(2 l+9) (2 l+13) }} &  \sqrt{ \frac{ 5(l+10)(l+9) }{ 6 (2 l+11) (2 l+13) }} \\[.5cm]
-\sqrt{   \frac{ 15  (l+2)(l+9) }{ 8 (2 l+9) (2 l+11)} }  &    -\frac{ (18-l)}{\sqrt{ 12(2 l+9) (2 l+13) }}        &  \sqrt{  \frac{ 49 (l+3)(l+10)}{ 24 (2 l+11) (2 l+13)}} \\[.5cm]
\sqrt{  \frac{5(l+9)(l+10)}{8(2 l+9) (2 l+11) }} 	     &   -  \sqrt{ \frac{ 9(l+2)(l+10) }{ 4(2 l+9) (2 l+13) }}& \sqrt{ \frac{ 9 (l+2)(l+3) }{ 8 (2 l+11) (2 l+13) }}
\end{array}
\right),
\end{array}
\eeq
and for odd spins, $\eta_i^{\rm odd}$ is consistent with $\eta_i^{\rm even}$ and the exchange symmetry $l\rightarrow -l- 12$, and correspondingly 
\bea
\widetilde{c}(4|l\ {\rm odd})&=&
\left(
\begin{array}{ccc}
\sqrt{  \frac{ 5 (l+2)(l+3) }{ 6 (11 + 2 l) (13 + 2 l) } }  &    \sqrt{ \frac{ 5 (l+3)(l+9)}{ 3(11 + 2 l) (15 + 2 l) }} &  \sqrt{ \frac{ 3(l+10)(l+9) }{ 2 (13 + 2 l) (15 + 2 l) }} \\[.5cm]
-\sqrt{  \frac{ 49(l+2)(l+9) }{ 24 (11 + 2 l) (13 + 2 l)} }  &    -\frac{ (l+30)}{\sqrt{ 12(11 + 2 l) (15 + 2 l) }}        &  \sqrt{  \frac{ 15(l+3)(l+10)}{ 8 (13 + 2 l) (15 + 2 l) }} \\[.5cm]
\sqrt{  \frac{ 9(l+9)(l+10)}{8(11 + 2 l) (13 + 2 l)}} 	     &     -\sqrt{ \frac{ 9(l+2)(l+10) }{ 4 (11 + 2 l) (15 + 2 l) }}&\sqrt{ \frac{ 5(l+2)(l+3) }{ 8 (13 + 2 l) (15 + 2 l)}}
\end{array}
\right).
\eea
At twist 11, we have found
\beq
\eta_{i}^{\rm even}=
\left\{
\begin{array}{l}
-\frac{2240 (l+8)(l+9)}{(l+1)(l+2)(l+3)(l+4)}, -\frac{2240(l+9)}{(l+3)(l+4)(l+7)}, -\frac{2240}{(l+7)(l+10)} , -\frac{2240(l+5)(l+6)}{(l+7)(l+10)(l+11)(l+12)}
\end{array}
\right\}.
\eeq
For higher twists the solution of $\widetilde{c}(t|l)$ becomes quite lengthy and it is helpful to introduce a more compact notation. We define
\beq
(n) = \sqrt{l+n}\,,\qquad [n] = \sqrt{2l + n}\,.
\eeq
Then,
\bea
\label{twist11ctilde}
\widetilde{c}(5|l\ {\rm even} )&=&\left(
\begin{array}{cccc}
\sqrt{\frac{33}{16}} \frac{(2) (3) (4)}{[9] [11] [13]} 	&   \sqrt{\frac{45}{16}} \frac{(3) (4) (10)}{[9] [13] [15]}     &    \sqrt{\frac{35}{16}} \frac{(4) (10) (11)}{[11] [13] [17]}   &  \sqrt{\frac{15}{16}} \frac{(10) (11) (12)}{[13] [15] [17]} \\[.3cm]
-\sqrt{\frac{55}{16}} \frac{(2) (3) (10)}{[9] [11] [13]} 	&   -\sqrt{\frac{3}{16}}  \frac{(l+32)(3)}{[9] [13] [15]}       &    -\sqrt{\frac{21}{16}}  \frac{(2-l)(11)}{[11] [13] [17]}       &    \sqrt{\frac{7}{16}} \frac{(4) (10) (12)}{[13] [15] [17]} \\[.3cm]
\sqrt{\frac{33}{16}} \frac{(2) (9) (11)}{[9] [11] [13]}   &   \sqrt{\frac{1}{80}} \frac{ (24-13l)(11)}{[9] [13] [15]}        &   -\sqrt{\frac{63}{80}}  \frac{(22+l)(3)}{[11] [13] [17]}      &   \sqrt{\frac{297}{80}} \frac{(3) (4) (12)}{[13] [15] [17]} \\[.3cm]
-\sqrt{\frac{7}{16}} \frac{(10) (11) (12)}{[9] [11] [13]} &   \sqrt{\frac{231}{80}} \frac{(2) (11) (12)}{[9] [13] [15]} &   -\sqrt{\frac{297}{80}} \frac{(2) (3) (12)}{[11] [13] [17]}  &     \sqrt{\frac{77}{80}} \frac{(2) (3) (4)}{[13] [15] [17]} \\[.3cm]
\end{array}
\right).\notag
\eea
The solution in the odd sector is given in the obvious way by the exchange symmetry $l\rightarrow -l -14$.

\subsubsection{General results}

The anomalous dimensions follow a simple pattern as we vary $t$, 
\bea
\eta_{i}^{\rm even}(t)\Big|_{i=1}^{ t-1  } &=& - \frac{2 \left(  t-1\right)_2 \left(  t+2\right)_2 \left(  l+t\right)_2\left(  l+t+3\right)_2 }{   \left( l+2i-1 \right)_{6} } \label{anom_even}\\[.3cm]
\eta_{i}^{\rm odd}(t)\Big|_{i=1}^{t-1 } &=& -\frac{2 \left(  t-1\right)_2 \left(  t+2\right)_2   \left(  l+t\right)_2\left(  l+t+3\right)_2 }{   \left( l+2i \right)_{6} } \label{anom_odd}
\eea
which has been explicitly confirmed up to $t=12$. Formulas \eqref{anom_even} and \eqref{anom_even} 
can be immediately compared with the anomalous dimensions of $K_{t,l,i}$ in the $[0,0,0]$ channel, 
which we rewrite in the form,
\beq
\eta_{t,l,i} = -\frac{2(t-1)_2(t+1)_2 (l+t)_{2}(l+t+2)_{2} }{(l+2i-1)_{6}}\,,\quad l\ {\rm even}
\label{singletanom}
\eeq
Experimentally, we see that going from $[0,0,0]$ to the $[0,1,0]$ representation can be accounted by introducing a gap in the Pochhammer structure. 
Looking instead at the structure of $\widetilde{c}(t|l)$ we obtain the following generalizations: 
For even spins, 
\begin{align}
\widetilde{c}(t|l\ {\rm even})_{p=2,\ldots t}^{i=1\ldots t-1}\notag\\=&
				\sqrt{ \frac{2^{1-t} (2 l++3+4i) \left((i+l+1)_{-i-p+t+1}\right){}^{\sigma_1}
								  	  \left((l+p+t+3)_{i-p+1}\right){}^{\sigma_2} }{\left(i+l+\frac{5}{2}\right)_{t-1} }  }\notag \\
								               & 
				\times \sum_{k=0}^{\min (i-1,p-2,-i+t-1,t-p)} l^k \a^{\rm even} (p-1,i,k)
\end{align}
For odd spins, 
\begin{align}
\widetilde{c}(t|l\ {\rm odd})_{p=2,\ldots t}^{i=1\ldots t-1}\notag\\= &
				\sqrt{\frac{2^{1-t} (2 l+5+4i) \left((i+l+1)_{-i-p+t+1}\right){}^{\sigma_1} 
									\left((l+p+t+3)_{i-p+1}\right){}^{\sigma_2}}{\left(i+l+\frac{7}{2}\right)_{t-1}}  }\notag \\
										&
				\times  \sum_{k=0}^{\min (i-1,p-2,-i+t-1,t-p)} l^k \a^{\rm odd}(p-1,i,k) .
\end{align}
The sign functions $\sigma_1$ and $\sigma_2$ in both case are given by, 
\beq
\sigma_1=\text{sgn}(-i-p+t+1),\qquad \sigma_2=\text{sgn}(i-p+1)
\eeq
Imposing orthonormality remarkably fix the unknown constants $\a^{\rm even}(p-1,i,k)$ and $\a^{\rm odd}(p-1,i,k)$ uniquely. Let us remark that the difference in the two cases only comes from the form of the denominators. 
For $p=2$, which is relevant to $C_{23;i}$, we have been able to find the following general formula, 
\begin{align}
	\a(1,i,0){}^2=\left\{
	\begin{array}{ll}
	\frac{2^{1-t} (2 i+2)! (t-2)! (2 t-2 i+3)!}{3 (i-1)! (i+1)! (t+3)! (t-i-1)! (t-i+1)!}\qquad &l\text{ even} \\[.2cm]
	\frac{2^{1-t} (2 i+3)! (t-2)! (2 t-2 i+2)!}{3 (i-1)! (i+1)! (t+3)! (t-i-1)! (t-i+1)!}\qquad &l\text{ odd}
	\end{array}
	\right.\ .
\end{align}
We thus have all the data we need in order to bootstrap the double discontinuity of $\langle \cO_2\cO_3\cO_2\cO_3\rangle$ at one loop.


\section{One loop}

\label{sec5}

We would now like to use the data we have obtained from the solution to the mixing problem considered in the previous section to bootstrap the order $a^2$ contributions to the correlators $\langle \mathcal{O}_2 \mathcal{O}_2 \mathcal{O}_3 \mathcal{O}_3\rangle$ and $\langle \mathcal{O}_2 \mathcal{O}_3 \mathcal{O}_2 \mathcal{O}_3\rangle$. We must first perform the summations which describe the double discontinuities and then try to construct the full function $F^{(2)}(u,v)$ (or equivalently $G^{(2)}(u,v)$) which has those discontinuities. 
\subsection{Constructing the double discontinuities}

Let us first consider the double discontinuity of $\langle \mathcal{O}_2 \mathcal{O}_2 \mathcal{O}_3 \mathcal{O}_3\rangle$, or equivalently the function $F^{(2)}_2(u,v)$ from the expansion (\ref{FGlogexp}). Let us also recall that the OPE predicts, 
\be\label{F2ddiscexp0}
F_2^{(2)}(u,v) =\frac{1}{2} (\log u)^2\sum_{t,l,j}\, \bigl(\eta^{(1)}_{t,l,i}\bigr)^2 \bigl(C^{(0)}_{22;K_{t,l,i}}C^{(0)}_{33; K_{t,l,i}}\bigr)  \LL^{2233}_{[0,0,0]}(t |l)
\ee
where the anomalous dimensions $\eta^{(1)}_{t,l,i}$, given in \eqref{anomn0n}, refer to the double trace operators $\{K_{t,l,i}\}_{i=1}^{t-1}$. 
In this case, the sum over $l$ runs only on even spins, namely $l=0,2,\ldots,\infty$.
The result for the OPE coefficients $C^{(0)}_{22;K_{t,l,i}}$ and $C^{(0)}_{33; K_{t,l,i}}$ can be obtained from \cite{unmixing} and we repeat it here for convenience:

\bea
C^{(0)}_{22;K_{t,l,i}}&=& + \sqrt{ \tfrac{8(t+l+1)!^2 t!^2 (l+1)(2t+l+2)}{(2t)!(2t+2l+2)!}\, \R^{22}_{t,l,i} \,  } \times \a^{22}_{t,i}\\[.2cm]
C^{(0)}_{33; K_{t,l,i}}&=&  \sqrt{ \tfrac{9(t+l+1)!^2 t!^2 (l+1)(2t+l+2)(t-2)(t+3)(l+t-1)(l+t+4)}{12(2t)!(2t+2l+2)!}\, \R^{33}_{t,l,i} \,  }\times \a^{33}_{t,i}  \qquad
\eea 
where
\footnote{In parametrizing $C^{(0)}_{33; K_{t,l,i}}$ we used slightly different conventions for $\R^{33}_{t,l,i}$ and $\a^{33}_{t,i}$ compared to \cite{unmixing}. } 
\bea
\R^{22}_{t,l,i} = \tfrac{2^{1-t}(2l+3+4i) (l+i+1)_{t-i-1} (t+l+4)_{i-1}}{\left(\frac{5}{2}+l+i\right)_{t-1}}, & 
\a^{22}_{t,i} = \sqrt{\tfrac{2^{(1 - t)} (2 + 2 i)! (t-2)! (2t - 2 i + 2)!}{3 (i-1)! (i+1)! (t+2)! (t-i-1)! (t-i+1)!}}\,, \\
\R^{33}_{t,l,i} = \tfrac{2^{1-t}(2l+3+4i) (l+i+1)^{ }_{t-i-2} (t+l+5)^{ }_{i-2}}{\left(\frac{5}{2}+l+i\right)_{t-1}}, & 
\a^{33}_{t,i}=  \sqrt{\tfrac{ 6   }{(t-2)(t+3)}} {\scriptstyle ((3 + 3 i + 2 i^2 - 2 t - t^2) + (2 i - t) l)} \,\a^{22}_{t,i}\,.
\eea
By using the explicit form of the long superblocks (\ref{flong}) we can rewrite (\ref{eq1_log2u}) as
\be
\label{F2ddiscexp}
F_2^{(2)}(u,v)  = \frac{1}{2}\sum_{t=2}^{\infty}  \sum_{l\, {\rm even} }  \sum_{i=1}^{t-1} \, \bigl(\eta^{(1)}_{t,l,i}\bigr)^2 \bigl(  C^{(0)}_{22;K_{t,l,i}} C^{(0)}_{33;K_{t,l,i}}\bigr)  \frac{1}{u^4} \CB^{\,2+t|l} \,,
\ee
where
\be
\label{cBform}
\CB^{\,t|l}=(-1)^l\, (x\bar{x})^t\ \frac{x^{l+1}\, \bF_{t+l }(x)\bF_{t-1}(\bar{x})- \bar{x}^{l+1}\ \bF_{t-1}(x)\bF_{t+l }(\bar{x}) }{x-\bar{x}}
\ee
and $\bF_{t}(x) = {}_2F_1\left( t ,t ,2t;x\right)$ for the case at hand.

Similarly we may consider the double discontinuity of $\langle \mathcal{O}_2 \mathcal{O}_3 \mathcal{O}_2 \mathcal{O}_3\rangle$. 
From the OPE analysis in (\ref{double_disc_2323}), we recall 
\be
\label{G2ddiscexp}
G_2^{(2)}(u,v) = \frac{1}{2} \sum_{l=0}^{\infty} \sum_{t=2}^{\infty} \sum_{i=1}^{t-1} \bigl(\eta^{(1)}_{t,l,j}\bigr)^2 \bigl(C^{(0)}_{23;\cK_{t,l,j}}\bigr)^2 \frac{1}{u^{\frac{9}{2} }} \mathcal{B}^{2+t +\frac{1}{2} |l}
\ee
where now the anomalous dimensions $\eta^{(1)}_{t,l,j}$, given in \eqref{anom_even}-\eqref{anom_odd}, 
refer to the operators in the $[0,1,0]$ channel, $\cK_{t,l,j}$, with both even and odd spins, (i.e not to be confused with $K_{t,l,i}$ in the singlet channel.)
The explicit expression for the three-point functions is
\be
(C^{(0)}_{23;\cK_{t,l,j}}\bigr)^2= \frac{(l+t+2)!^2 (t+1)!^2 (l+1)(l+2t+3)}{2(2t)!(2t+2l+3)!}\, \R^{23}_{t,l,i} \a^{23}_{t,i} 
\ee
where for even spins
\be
\R^{23}_{t,l,i} =  \tfrac{ 2^{1-t} (2l+3+4i)(l+1+i)_{t-i-1} (l+t+5)_{i-1} }{ \left(\tfrac{5}{2}+l+i\right)_{t-1} },\qquad \a^{23}_{t,l,i}= \tfrac{2^{1-t} (2 i+2)! (t-2)! (2 t-2 i+3)!}{3 (i-1)! (i+1)! (t+3)! (t-i-1)! (t-i+1)!}
\ee
whereas for odd spins the values of $\R^{23}_{t,l,i}$ and $\a^{23}_{t,l,i}$ can be obtained upon using the symmetry, $i\rightarrow t-i$ and $l\rightarrow -l-2t-4$.
The conformal block $\mathcal{B}^{\,t|l}$ is again given by (\ref{cBform}) but now $\bF_{t}(x)$ has non trivial dependence on the external charges, i.e. $ \bF_{t}(x)= {}_2F_1\left( t +\tfrac{1}{2} ,t - \tfrac{1}{2} ,2t;x\right)$.

In order to perform the sums in (\ref{F2ddiscexp}) and (\ref{G2ddiscexp}) it is very useful to consider the action of certain Casimir operators, related to those considered in \cite{Alday:2016njk} which simplify the sums considerably. First we introduce the second order operators,
\begin{align}
D &= x^2 \partial_x (1-x) \partial_x -abx-(a+b)x^2 \partial_x,\\
\bar{D} &= \bar{x}^2 \partial_{\bar x} (1-\bar{x}) \partial_{\bar{x}} - ab\bar{x}-(a+b)\bar{x}^2 \partial_{\bar{x}},
\end{align}
where $a=-\frac{1}{2}\Delta_{12}$ and $b=\frac{1}{2}\Delta_{34}$.
From these operators we construct the Casimirs,
\begin{align} 
\mathcal{D}_2 &= D + \bar{D} + 2 \frac{x \bar{x}}{x-\bar{x}} \bigl((1-x)\partial_x - (1-\bar{x})\partial_{\bar{x}}\bigr)\,, \\
\mathcal{D}_4 &= \biggl(\frac{x \bar{x}}{x-\bar{x}}\biggr)^2 (D-\bar{D}) \biggl(\frac{x \bar{x}}{x-\bar{x}}\biggr)^{-2} (D - \bar{D})\,.
\label{conformalcasimirs}
\end{align}
On the conformal blocks $\mathcal{B}^{t+2|l}$ these operators have the following eigenvalues,
\begin{align}
\lambda_2 &= \frac{1}{2}\bigl(l(l+2) + (\tau + l)(\tau+l-4)\bigr)\,,\notag\\
\lambda_4 &= l(l+2)(\tau + l -1)(\tau +l -3)\,,
\label{eigenvalues}
\end{align}
where $\tau$ is even ($\tau = 2t+4$) in the $[0,0,0]$ channel and odd ($\tau = 2t+5$) in the $[0,1,0]$ channel.
Finally we construct the Casimir combinations
\begin{align}
\Delta^{(4)}_+ &= \mathcal{D}_4 - \mathcal{D}_2^2 + c \mathcal{D}_2 + d \, \\
\Delta^{(4)}_- &= \mathcal{D}_4 - \mathcal{D}_2^2 + e \mathcal{D}_2 + f \,,
\end{align}
where the constants $c,d,e,f$ depend on the OPE channel we are considering. For the cases at hand we need the $[0,0,0]$ channel for the $\langle \mathcal{O}_2 \mathcal{O}_2 \mathcal{O}_3 \mathcal{O}_3 \rangle$ correlator (and also for the $\langle \mathcal{O}_2 \mathcal{O}_2 \mathcal{O}_2 \mathcal{O}_2 \rangle$ correlator considered in \cite{Aprile:2017bgs}) and the $[0,1,0]$ channel for the $\langle \mathcal{O}_2 \mathcal{O}_3 \mathcal{O}_2 \mathcal{O}_3 \rangle$ correlator. In these cases we have
\begin{alignat}{5}
[0,0,0] &: \qquad &&c=6\,, \quad &&d=0\,,\quad  &&e=-2\,,\quad &&f=0\,,\\
[0,1,0] &: \qquad &&c=13\,, \quad &&d=-\frac{105}{4}\,,\quad  &&e=1\,,\quad &&f=\frac{15}{4}\,.
\end{alignat}

If we now consider the product
\be
\label{defDelta8}
\Delta^{(8)} = \Delta^{(4)}_+ \Delta^{(4)}_- 
\ee
we find that it factorises into holomorphic and antiholomorphic parts,
\be
\Delta^{(8)}  =  (x-\bar{x})^{-1} \mathcal{D}^{(2)}_+ \mathcal{D}^{(2)}_- \bar{\mathcal{D}}^{(2)}_+ \bar{\mathcal{D}}^{(2)}_- (x-\bar{x})\,,
\ee
where in the two channels under consideration here we have
\begin{alignat}{3}
[0,0,0]&:\qquad \mathcal{D}^{(2)}_- &&= -2x^3\partial_x (1-x) \partial_x x^{-1}\,, \qquad &&\mathcal{D}^{(2)}_+ = \mathcal{D}^{(2)}_- + 4\,,\notag \\
[0,1,0]&:\qquad \mathcal{D}^{(2)}_- &&= \frac{1}{2}(5-2x\partial_x)(1-x)(1-2x\partial_x)\,, \qquad &&\mathcal{D}^{(2)}_+ = \mathcal{D}^{(2)}_- - 6\,.
\end{alignat}
Moreover we find that $\Delta^{(8)}$ has the following eigenvalue in each channel
\begin{alignat}{2}
[0,0,0] &:\qquad  \lambda_8 &&= 16 (t-1)_2(t+1)_2 (l+t)_{2}(l+t+2)_{2}\\
[0,1,0] &:\qquad  \lambda_8 &&= 16 (t-1)_2 (  t+2)_2 (  l+t)_2(  l+t+3)_2\,.
\end{alignat}
Up to a factor of $(-8)$ these reproduce exactly the numerators of the anomalous dimensions given in (\ref{singletanom}) for the $[0,0,0]$ channel and (\ref{anom_even}), (\ref{anom_odd}) for the $[0,1,0]$ channel. This suggests that the sum may simplify if one pulls out factors made from the operator $\Delta^{(8)}$. Indeed we find this is the case. More precisely, for the leading discontinuities $F_2^{(2)}$ and $G_2^{(2)}$, one should pull out the operator
\be
-\frac{1}{8}\tilde{\Delta}^{(8)} = u^{-q} \Delta^{(8)} u^q
\ee
where $q=4$ for the $[0,0,0]$ channel and $q=\tfrac{9}{2}$ for the $[0,1,0]$ channel. This leads us to simple explicit results for the leading discontinuities.

For the first correlator of relevance here, $\langle \mathcal{O}_2 \mathcal{O}_2 \mathcal{O}_3 \mathcal{O}_3 \rangle$, we find the result for $F_2^{(2)}$ takes the form
\begin{align}
\label{F2ddiscresum}
F_2^{(2)}(u,v) &= -\frac{1}{8} \tilde{\Delta}^{(8)}_{[0,0,0]}\biggl[ \hat{p}(u,v) \frac{{\rm Li}_1(x)^2 - {\rm Li}_1(\bar{x})^2}{(x-\bar{x})^9} + 2\biggl[\hat{p}(u,v) + v^5 \hat{p}\biggl(\frac{u}{v},\frac{1}{v}\biggr)\biggr]\frac{{\rm Li}_2(x) - {\rm Li}_2(\bar{x})}{(x-\bar{x})^9} \notag \\
& \qquad + \hat{q}(u,v)\frac{{\rm Li}_1(x) + {\rm Li}_1(\bar{x})}{(x-\bar{x})^8} + \hat{r}(u,v)\frac{{\rm Li}_1(x) - {\rm Li}_1(\bar{x})}{(x-\bar{x})^9} + \frac{\hat{s}(u,v)}{(x-\bar{x})^8}\biggr]\,.
\end{align}
The coefficients above are given by the following expressions,
\begin{align}
\hat{p}(u,v) =& \frac{3}{2} u v^2 \bigl[3 u^2+u v-6 u-4 v^2+v+3\bigr]\,, \notag \\
\hat{q}(u,v) =& \frac{v-1}{8 u}\bigl[5 u^4  + 62 u^3 (1 + v) - 
6 u^2 (9 + 22 v + 9 v^2) - 14 u (1 - v)^2 (1 + v) + (1 - v)^4 \bigr]\,, \notag \\
\hat{r}(u,v) =& \frac{1}{8u}\bigl[4 u^6  - 21 u^5 (1 + v)   - 3 u^4 (11 - 42 v + 11 v^2) + 
4 u^3 (1 + v) (25 - 49 v + 25 v^2) \notag \\
&\qquad  - 
12 u^2 (1 - v)^2 (3 + v) (1 + 3 v) - 15 u (1 - v)^4 (1 + v) + (1 - v)^6 \bigr]\,, \notag \\
\hat{s}(u,v) =& \frac{1}{4}\bigl[9 u^4  - 43 u^3 (1 + v) - 
4 u^2 (6 - 41 v + 6 v^2) + 57 u (1 - v)^2 (1 + v)  + (1 - v)^4 \bigr]\,.
\end{align}
We remark that the double discontinuity of the correlator $\langle \mathcal{O}_2 \mathcal{O}_2 \mathcal{O}_2 \mathcal{O}_2 \rangle$ computed in \cite{Aprile:2017bgs} can similarly be simplified by using the same singlet channel $\Delta^{(8)}$ operator.

For the other correlator, $\langle \mathcal{O}_2 \mathcal{O}_3 \mathcal{O}_2 \mathcal{O}_3 \rangle$, the double discontinuity is given by
\begin{align}
\label{G2ddiscresum}
G_2^{(2)}(u,v) &=  -\frac{1}{8}\tilde{\Delta}^{(8)}_{[0,1,0]}\biggl[p(u,v) \frac{{\rm Li}_1(x)^2 - {\rm Li}_1(\bar{x})^2}{(x-\bar{x})^9} + \tilde{p}(u,v) \frac{{\rm Li}_2(x) - {\rm Li}_2(\bar{x})}{(x-\bar{x})^9} \notag \\
& \qquad + q(u,v)\frac{{\rm Li}_1(x) + {\rm Li}_1(\bar{x})}{(x-\bar{x})^8} + r(u,v)\frac{{\rm Li}_1(x) - {\rm Li}_1(\bar{x})}{(x-\bar{x})^9} + \frac{s(u,v)}{(x-\bar{x})^8}\biggr]\,.
\end{align}
The coefficients above are given by
\begin{align}
p(u,v)=&-\frac{3v^2}{2}\bigl[3 - 6 u + 3 u^2 - 6 v + 8 u v + 3 v^2\bigr]\,,\notag\\
\tilde{p}(u,v)=&-3 \bigl[u^3 + 3 u^2 (1 - v + v^2) - 
u (1 - v) (2 + 5 v + 8 v^2) - (1 - v)^3 (2 + 3 v)  \bigr] \,, \notag \\
q(u,v) =&\frac{1}{8u}\bigl[u^4  + 2 u^3 (15 - 4 v) +
2 u^2 (9 - 24 v - 37 v^2)  - 2 u (1 - v)^2 (23 + 42 v) - 3 (1 - v)^4 \bigr]\,, \notag \\
r(u,v) =&-\frac{3}{8u}\bigl[u^5 - u^4 (15 - v) + 2 u^3 (6 + 19 v - 19 v^2) + 2 u^2 (1 - v) (8 - 23 v - 7 v^2) \notag \\
&\qquad    - u (1 - v)^3 (13 + 21 v) - (1 - v)^5\bigr]\,, \notag \\
s(u,v) =&-\frac{1}{8} \bigl[ u^4  + 2 u^3 (3 - 2 v) -
6 u^2 (21 - 2 v - v^2) \notag \\
&\qquad + 2 u (1 - v) (31 + 149 v + 2 v^2) + (57 - v) (1 - v)^3 \bigr]\,.
\end{align}

Given the explicit forms of the $\Delta^{(8)}$ operators in each case we may simply compute the full result for $F_2^{(2)}$ and $G_2^{(2)}$. They take the form,
\begin{align}
\label{F2ddiscresum}
F_2^{(2)}(u,v) &=  {\hat P}(u,v) \frac{{\rm Li}_1(x)^2 - {\rm Li}_1(\bar{x})^2}{x-\bar{x}} + 2\biggl[ \frac{1}{v^3} {\hat P}\biggl(\frac{u}{v},\frac{1}{v}\biggr)+ {\hat P}(u,v) \biggr]\frac{{\rm Li}_2(x) - {\rm Li}_2(\bar{x})}{x-\bar{x}} \notag \\
& \qquad + {\hat Q}(u,v)\bigl({\rm Li}_1(x) + {\rm Li}_1(\bar{x})\bigr) + {\hat R}(u,v)\frac{{\rm Li}_1(x) - {\rm Li}_1(\bar{x})}{x-\bar{x}} + {\hat S}(u,v)\,.
\end{align}
and
\begin{align}
\label{G2ddiscresum}
G_2^{(2)}(u,v) &=  P(u,v) \frac{{\rm Li}_1(x)^2 - {\rm Li}_1(\bar{x})^2}{x-\bar{x}} + 2\biggl[ \frac{1}{v^3}{\hat P}\biggl(\frac{1}{v},\frac{u}{v}\biggr)+P(u,v)\biggr]\frac{{\rm Li}_2(x) - {\rm Li}_2(\bar{x})}{x-\bar{x}} \notag \\
& \qquad + Q(u,v)\bigl({\rm Li}_1(x) + {\rm Li}_1(\bar{x})\bigr) + R(u,v)\frac{{\rm Li}_1(x) - {\rm Li}_1(\bar{x})}{x-\bar{x}} + S(u,v)\,.
\end{align}

The coefficient functions $P,Q,R,S$ and similarly the hatted quantities are rational functions of $x$ and $\bar{x}$ with denominators of the form $(x-\bar{x})^{16}$, and are symmetric under $x \leftrightarrow \bar{x}$.  Note that the symmetry of the full correlation function, $G^{(2)}(u,v)=G^{(2)}(v,u)$, is visible in the double discontinuity $G_2^{(2)}$ for the term proportional to $\log^2 u\log^2 v$. Indeed we can verify that $P(v,u)=P(u,v)$.  On the other hand, we are able to express the coefficient function of ${\rm Li}_2$ in terms of ${\hat P}(u,v)$ and $P(1/v,u/v)$. This non trivial fact will be important when we will uplift the double discontinuity to a full correlation function.

\subsection{Uplifting to the full function}

The structure of the double discontinuities (\ref{F2ddiscresum}) and (\ref{G2ddiscresum}) is very similar to the double discontinuity found in \cite{Aprile:2017bgs} for the correlator $\langle \mathcal{O}_2 \mathcal{O}_2 \mathcal{O}_2 \mathcal{O}_2 \rangle$. This suggests that the transcendental functions appearing in the full one-loop contributions of $\langle \mathcal{O}_2 \mathcal{O}_2 \mathcal{O}_3 \mathcal{O}_3 \rangle$ and $\langle \mathcal{O}_2 \mathcal{O}_3 \mathcal{O}_2 \mathcal{O}_3 \rangle$ will also be given by the same one-loop and two-loop ladder functions which arise in the case of $\langle \mathcal{O}_2 \mathcal{O}_2 \mathcal{O}_2 \mathcal{O}_2 \rangle$. We recall that they take the form \cite{Usyukina:1992jd},
\be
\Phi^{(l)}(u,v) = - \frac{1}{x-\bar{x}} \phi^{(l)}\biggl(\frac{x}{x-1},\frac{\bar{x}}{\bar{x}-1}\biggr)\,,
\ee
where the pure transcendental part is given by
\be
\phi^{(l)}(x,\bar{x}) = \sum_{r=0}^l (-1)^r \frac{(2l-r)!}{r!(l-r)! l!} \log^r (x \bar{x}) ({\rm Li}_{2l-r}(x) - {\rm Li}_{2l-r}(\bar{x}))\,.
\ee
We recall also the crossing symmetry of the ladder functions,
\be
\phi^{(l)}\biggl(\frac{1}{x},\frac{1}{\bar{x}}\biggr) = - \phi^{(l)}(x,\bar{x})\,.
\ee
The one-loop function also obeys
\be
\phi^{(1)}(1-x,1-\bar{x}) = - \phi^{(1)}(x,\bar{x})\,.
\ee

We proceed very much as in the case of the correlator $\langle \mathcal{O}_2 \mathcal{O}_2 \mathcal{O}_2 \mathcal{O}_2 \rangle$ investigated in \cite{Aprile:2017bgs}. We make an ansatz for $F^{(2)}(u,v)$ (or equivalently $G^{(2)}(u,v)$) in terms of single-valued harmonic polylogarithms with coefficients wich are rational functions of $x$ and $\bar{x}$ with denominators of the form $(x-\bar{x})^{17}$, to match the double discontinuities (\ref{F2ddiscresum}) and (\ref{G2ddiscresum}). We demand that our ansatz reproduces correctly both double discontinuities and furthermore that the resulting function does not have any poles at $x=\bar{x}$. This set of constraints produces a particular solution with four free parameters. To express the dependence we first quote the  particular solutions, $G_p^{(2)}$ and $F_p^{(2)}$, and then describe the four remaining degrees of freedom. For convenience we quote first the form of $G_p^{(2)}(u,v)$,
\begin{align}
\label{G2uvresult}
G_p^{(2)}(u,v) = &A_1(x,\bar{x}) \phi^{(2)}(x',\bar{x}') + A_2(x,\bar{x}) \phi^{(2)}(x,\bar{x}) + A_2(1-x,1-\bar{x})\phi^{(2)}(1-x,1-\bar{x}) \notag \\
& + \bigl[A_3(x,\bar{x}) x(1-x)\partial_x \phi^{(2)}(x',\bar{x}') + (x \leftrightarrow \bar{x})\bigr] \notag \\
& + \bigl[A_4(x,\bar{x}) x \partial_x \phi^{(2)}(x,\bar{x}) +(x \leftrightarrow \bar{x})\bigr] \notag \\
& - \bigl[A_4(1-x,1-\bar{x}) (1-x) \partial_x \phi^{(2)}(1-x,1-\bar{x}) +(x \leftrightarrow \bar{x})\bigr] \notag\\
& + A_5(x,\bar{x})\log^2(u/v)  + A_6(x,\bar{x}) \log^2 u + A_6(1-x,1-\bar{x})\log^2 v \notag \\
&+ A_7(x,\bar{x}) \phi^{(1)}(x,\bar{x}) + A_8(x,\bar{x})  \log u + A_8(1-x,1-\bar{x}) \log v  + A_9(x,\bar{x})\,,
\end{align}
where we have used the notation $x' = \frac{x}{x-1}$. The explicit expressions for the coefficient functions $A_1,\ldots,A_9$ are rather cumbersome but we provide them in a {\tt Mathematica} notebook attached to the arXiv submission of this article. These functions obey,
\begin{align}
\begin{array}{llllll}
A_1(\bar{x},x) &=& -A_1(x,\bar{x})\,,\qquad &A_1(1-x,1-\bar{x}) &=&  - A_1(x,\bar{x})\,,\notag \\
A_2(\bar{x},x) &=& -A_2(x,\bar{x})\,, &&\notag \\
A_3(\bar{x},x) &=& +A_3(x,\bar{x})\,, \qquad &A_3(1-x,1-\bar{x}) &=& +A_3(x,\bar{x})\,,\notag \\
A_5(\bar{x},x) &=& +A_5(x,\bar{x})\,, \qquad &A_5(1-x,1-\bar{x}) &=& +A_5(x,\bar{x})\,,\notag \\
A_6(\bar{x},x) &=& +A_6(x,\bar{x})\,, \notag \\
A_7(\bar{x},x) &=& -A_7(x,\bar{x})\,, \notag \qquad &A_7(1-x,1-\bar{x}) &=& -A_7(x,\bar{x})\,,\\
A_8(\bar{x},x) &=& +A_8(x,\bar{x})\,, \notag \\
A_9(\bar{x},x) &=& +A_9(x,\bar{x})\,, \qquad &A_9(1-x,1-\bar{x}) &=& +A_9(x,\bar{x})\,.
\end{array}
\end{align}
The properties above are necessary for $G_p^{(2)}(u,v)$ to be symmetric under $x \leftrightarrow \bar{x}$ and for the crossing property $G_p^{(2)}(v,u)=G_p^{(2)}(u,v)$ to hold.
Part of the weight four function in the first line of \eqref{G2uvresult} can be immediately related to $G_{2}^{(2)}$. In particular, we recognize
\beq
\frac{P(u,v)}{x-\bar{x}}=-\frac{ A_1(x,\bar{x}) }{4},\qquad   \frac{1}{v^3}{\hat P}\biggl(\frac{1}{v},\frac{u}{v}\biggr)+P(u,v)  = -\frac{ A_1(x,\bar{x})  }{4} +\frac{A_2(x,\bar{x}) }{4}
\eeq
whereas the remaining coefficient functions $q(u,v)$ $r(u,v)$ and $s(u,v)$ enter non trivially into the set of $A_i(x,\bar{x})$. 

The particular solution $F_p^{(2)}(u,v)$ is given by applying the crossing transformation \eqref{FGrel}, $F(u,v) = {1}/{u^4} G\left( {1}/{u},{v}/{u}\right)$, to the function \eqref{G2uvresult},
\begin{align}
F_p^{(2)}(u,v) = \frac{1}{u^4}\Bigl[&-\hat{A}_2(x',\bar{x}') \phi^{(2)}(x',\bar{x}') - \hat{A}_2(x,\bar{x}) \phi^{(2)}(x,\bar{x}) - \hat{A}_1(x,\bar{x})\phi^{(2)}(1-x,1-\bar{x}) \notag \\
&+  \bigl[\hat{A}_4(x',\bar{x}') x(1-x) \partial_x \phi^{(2)}(x',\bar{x}') + (x \leftrightarrow \bar{x})\bigr] \notag\\
& + \bigl[\hat{A}_4(x,\bar{x}) x \partial_x \phi^{(2)}(x,\bar{x}) +(x \leftrightarrow \bar{x})\bigr] \notag \\
& - \bigl[\hat{A}_3(x,\bar{x}) (1-x)\partial_x \phi^{(2)}(1-x,1-\bar{x}) + (x \leftrightarrow \bar{x})\bigr] \notag \\
& + \hat{A}_6(x',\bar{x}')\log^2(u/v)  + \hat{A}_6(x,\bar{x}) \log^2 u + \hat{A}_5(x,\bar{x})\log^2 v  \notag \\
& -\hat{A}_7(x,\bar{x}) \phi^{(1)}(x,\bar{x}) - \bigl[\hat{A}_8(x,\bar{x}) + \hat{A}_8(x',\bar{x}') \bigr]  \log u  \notag \\
&+ \hat{A}_8(x',\bar{x}') \log v  + \hat{A}_9(x,\bar{x})\Bigr]\,,
\end{align}
where the functions $\hat{A}_1,\ldots, \hat{A}_9$ are related to $A_1,\ldots,A_9$ via $\hat{A}_i(x,\bar{x}) = A_i(1/x,1/\bar{x})$.

Now let us describe the four ambiguities. We find that they can be described in terms of the following four $\bar{D}$-functions,
\be
G^{(2)}(u,v) = G^{(2)}_p(u,v) + \alpha \overline{D}_{4444}(u,v) + \beta \overline{D}_{4545}(u,v) + \gamma \overline{D}_{4646}(u,v) + \delta v \overline{D}_{4565}(u,v)\,.
\ee


\subsection{Twist 4 sector of $\langle 2233 \rangle$ }

Within our ansatz we have obtained a one loop solution for $\langle \cO_2\cO_2\cO_3\cO_3\rangle$ with 4 free coefficients. 
Can we further constrain these coefficients? The answer is affirmative, and in fact there are further consistency 
conditions that our one-loop result must satisfy: Consider the expansion of the correlator up to second order in $a$, 
\begin{align}
\label{check_2233}
\langle \cO_2\cO_2\cO_3\cO_3\rangle = A \Bigl[ \langle \cO_2\cO_2\cO_3\cO_3\rangle_{\rm free}^{(0)} &+ a \langle \cO_2\cO_2\cO_3\cO_3\rangle_{\rm free}^{(1)} \notag \\
&+ a \langle \cO_2\cO_2\cO_3\cO_3\rangle^{(1)}_{\rm int} + a^2 \langle \cO_2\cO_2\cO_3\cO_3\rangle^{(2)}_{\rm int}+ \ldots  \Bigr]\,.
\end{align}
Above we have highlighted the fact that some contributions are the same as in free theory.
In the long sector there exists a single twist 4  double trace operator $K_{2,l}\sim \cO_2 \partial^l \cO_2$.
The 3-point function $\langle \cO_3\cO_3 K_{2,l}\rangle$ vanishes at leading order, and therefore
\beq
C_{22;K_{2,l}} C_{33; K_{2,l} }= A^{2233}(2|l) + a B^{2233}(2|l)+O(a^2),
\eeq
starts at subleading order, i.e.  $A^{2233}(2|l)=0$. Then, it follows from the OPE that, 
\begin{align}
\Bigl[\langle \cO_2\cO_2\cO_3\cO_3\rangle_{\rm free}^{(1)} + &\langle \cO_2\cO_2\cO_3\cO_3\rangle^{(1)}_{\text{int}} \Bigr]_{\rm twist \,4 \,long}=  \sum_{l} \symbolB^{2233}(2|l)\, \LL_{[0,0,0]}^{2|l}\\  
&\langle \cO_2\cO_2\cO_3\cO_3\rangle^{(2)}_{\rm int}\Big|_{\rm twist \,4 \,long}= \log\,u \sum_{l} \eta_{K_{2,l}} \symbolB^{2233}(2|l)\, \LL_{[0,0,0]}^{2|l} +\ldots 
\end{align}
where $\eta_{K_{2,l}}=-48/(l+1)(l+6)$ has been computed in  \cite{unmixing} and the dots stand for terms  analytic at $u=0$ which are not relevant here. 
The coefficients $B^{2233}(2|l)$ can be obtained
from $u^3\Db^{\rm \, sing}_{3522}$ of the corresponding supergravity amplitude, and are given by
\beq
B^{2233}(2|l)= 240N^3 \frac{ ((l+3)!)^2}{(2 l+6)!}\ .\label{c222c332}
\eeq
Thus the twist 4 sector of the $\log u$ part of the one-loop correlator is fully determined by the knowledge of 
\eqref{c222c332} and the anomalous dimension, $\eta_{K_{2,l}}$.  
It is interesting to notice in \eqref{c222c332} that the contributions from free theory and supergravity have
 the same $l$ dependence but differ in the overall coefficient, $24N^3$ and $216N^3$, respectively. 
Very nicely we find that this OPE constraint is consistent with our one-loop result and fixes two of the four remaining constants, namely
\begin{align}
	\alpha=0, \qquad \delta=0\,.
\end{align}   
We thus have  a solution with 2 remaining free parameters.


\section{Twist 5 anomalous dimensions at one-loop}
\label{sec6}

We now extract twist 5 anomalous dimensions from our one loop correlator $\langle \cO_2\cO_3\cO_2\cO_3\rangle$. We focus on $\cK_{2,l}$ because this is the only case in which there is a single operator for each spin, and we thus have enough information to determine its anomalous dimension. For higher twist there is a higher order  mixing problem to undo, and we expect further mixing with triple-trace operators to spoil predictability. 
The expansion of the correlator up to order $a^2$ takes the form
\begin{align}
\label{check_2323}
\langle \cO_2\cO_3\cO_2\cO_3\rangle = A \Bigl[ \langle \cO_2\cO_3\cO_2\cO_3\rangle_{\rm free}^{(0)} &+ a \langle \cO_2\cO_3\cO_2\cO_3\rangle_{\rm free}^{(1)} \notag \\
&+ a \langle \cO_2\cO_3\cO_2\cO_3\rangle^{(1)}_{\rm int} + a^2 \langle \cO_2\cO_3\cO_2\cO_3\rangle^{(2)}_{\rm int}+ \ldots  \Bigr]\,,
\end{align}
where 
\bea
&\langle \cO_2\cO_3\cO_2\cO_3\rangle_{\text{free}}^{(0)} =& +6\, \d_{13}^2 \d_{24}^3\\
&\langle \cO_2\cO_3\cO_2\cO_3\rangle_{\text{free}}^{(1)}=& +36 \left(\d_{13}\d_{24}^2\d_{12}\d_{34}+\d_{13}\d_{24}^2\d_{14}\d_{23}+2 \d_{24}\d_{12}\d_{34}\d_{14}\d_{23}\right)\\
&\langle \cO_2\cO_3\cO_2\cO_3\rangle^{(1)}_{\text{int}} =& -36\, \d_{12}^2 \d_{34}^2 \d_{24} \frac{ s(x,\bar{x},y,\bar{y})}{(y\bar{y})^2}\frac{u^2}{v}\Db_{5,2,3,2}
\eea
and our new result, given in~\eqref{G2uvresult}, is 
\begin{align}
\langle \cO_2\cO_3\cO_2\cO_3\rangle^{(2)}_{\text{int}} =  
g_{12}^2 g_{34}^2 g_{24} \, \mathcal{I}(u,v;\sigma,\tau) \, u^2 G^{(2)}(u,v)\,,
\end{align}
In order to extract twist 5 anomalous dimensions it is enough to restrict ourselves to the superconformal partial wave expansion in the long sector. 
The OPE at twist 5 implies the identity, 
\bea
&&
\langle \cO_2\cO_3\cO_2\cO_3  \rangle\Big|_{\text{twist 5}} =  \\
				&& \rule{.5cm}{0pt}
				A \d_{12}^2 \d_{34}^2 \d_{24}  \Bigg(
				\sum_l A_{2,l} \ \LL_{[0,1,0]}^{2323}(\tfrac{5}{2}|l) + \label{t4} \notag\\
			        &&\rule{1.9cm}{0pt}
			        \ a\,\sum_l \, \left[ \left( A_{2,l}\, \eta^{(1)}_{2,l} \log u + B_l \right)\LL_{[0,1,0]}^{2323}(\tfrac{5}{2}|l) 
															+ A_{2,l}\, \eta^{(1)}_{2,l} \partial_t\LL_{[0,1,0]}^{2323}(\tfrac{5}{2}|l) \right]+ \notag\\
			        &&\rule{1.9cm}{0pt}
			        \ a^2\sum_l \Big[ \tfrac{1}{2} A_{2,l}\big(\eta^{(1)}_{2,l}\big)^2 \log^2 u\ \LL_{[0,1,0]}^{2323}(\tfrac{5}{2}|l)  + \notag\\
				&&\rule{3cm}{0pt}
				 \left( A_{2,l}\, \eta^{(2)}_{2,l} + B_l\eta^{(1)}_l \right)\log u\ \LL_{[0,1,0]}^{2323}(\tfrac{5}{2}|l) + 2A_{2,l} \big(\eta^{(1)}_{2,l}\big)^2 \partial_t\LL_{[0,1,0]}^{2323}(\tfrac{5}{2}|l) +\ldots\Big] 
				 \rule{1cm}{0pt}\notag\\
				 \label{twist5OPE2323}
\eea
where $\eta_{2,l}^{(i=1,2)}\equiv\eta^{(i=1,2)}_{\cK_{2,l}}$ and we defined, 
\beq
C_{23;\cK_{2,l}} C_{23; \cK_{2,l} }= A_{2,l} + a B_{2,l}+O(a^2). 
\eeq
Therefore, we also have $A_{2,l}=\cA^{\{ 2323\} }(2|l)$, and $A_{2,l} \eta^{(1)}_{2,l}=\symbol^{\{2323\}}(2|l)$, which are known from \eqref{App+1pp+1} and \eqref{M2323}, respectively. 
We repeat them for convenience:
\beq
A_{2,l}=\frac{9 (l+1) (l+7) ((l+4)!)^2}{10 (2 l+7)!},\qquad 
\eta^{(1)}_{2,l} =\left\{ \begin{array}{ll} -\frac{80}{(l+1) (l+4)} \quad &l \text{ even}\\[.2cm] -\frac{80}{(l+4) (l+7)} \quad &l \text{ odd}\end{array}\right.
\eeq
We will now equate the expansion \eqref{twist5OPE2323} with a superconformal block expansion of the correlator in the long sector, and
determine the one loop correction to the anomalous dimension, $\eta_l^{(2)}$, from the last line. 

We proceed by first computing the coefficients $B_{2,l}$.  
It is convenient to separate the contribution to $B_{2,l}$ in free theory and tree level supergravity.
The conformal partial wave analysis of the free theory gives
\begin{align}
B_{2,l;\, \text{free}}= \frac{36 ((l+4)!)^2}{5 (2 l+8)!} \left\{ \begin{array}{l} (l+7)\quad l\ {\rm even}\\[.2cm] (l+1)\quad l\ {\rm odd} \end{array}\right. \ .
\end{align} 
Next, we can expand both $\partial_t\LL_{[0,1,0]}^{2323}(\tfrac{5}{2}|l)$ and $\Db^{\rm\, analytic}_{5332}$ in superblocks, and keep only the coefficients at twist 5. 
We then obtain values for the correction to the normalisation due to supergravity, $B_{\text{int},l}$, which together with the free theory yields the result
\begin{align}
B_{2,l}=B_{2,l;\,\text{free}}+B_{2,l;\text{int}}=\left\{
\begin{array}{ll}
\frac{((l+3)!)^2 \left(144 (l+4) (l+7) \left(H_{l+3}-H_{2 l+7}\right)+\frac{54}{5} \left(7 l^2+97l+296\right)\right)}{ (2 l+7)!}  \qquad &l \text{ even} \notag\\[.2cm]
\frac{((l+3)!)^2 \left(144 (l+1) (l+4) \left(H_{l+3}-H_{2 l+7}\right)+\frac{18}{5} (l+1) (21 l+104)\right)}{ (2l+7)!} \qquad &l \text{ odd}
\end{array}
\right.
  \end{align}
 consistent with the relation~\cite{0907.0151,Fitzpatrick:2011dm}
 \begin{align}
 	B_{2,l}=\frac \partial{\partial t} \symbol^{\{2323\}}(t|l)\Big|_{t=2}\label{ada}
 \end{align}
Note that in interpreting~\eqref{ada} we used the full $t$ dependence of $\symbol^{\{2323\}}$ and treated the even and odd spin formulae as completely independent formulae.
We can finally consider the one loop result at order $a^2$. The coefficient multypling $\partial_t\LL_{[0,1,0]}^{2323}(\tfrac{5}{2}|l)$ is 
\begin{align}
A_{2,l}\big(\eta^{(1)}_l\big)^2=\frac{5760 ((l+3)!)^2}{(2 l+7)!}
\left\{ \begin{array}{ll} 
\frac{l+7}{l+1}\qquad & l \text{ even}\\[5pt]
\frac{l+1}{l+7}\qquad & l \text{ odd}
\end{array}
\right.
\end{align}
Rearranging, we obtain $\eta^{(2)}_\ell$ directly 

\begin{align}
\eta^{(2)}_{2,l} = \left\{ \begin{array}{ll}
 \frac{  320 \left(9 l^4+68 l^3-1151 l^2-5738 l-3688\right)}{(l-1) (l+1)^3 (l+4)^3 (l+8)} \qquad &l=2,4,\dots\\[5pt]
 \frac{ 320 \left(9 l^4+140 l^3-487 l^2-11262 l-29400\right)}{l (l+4)^3 (l+7)^3 (l+9)}\qquad &\l=3,5,\dots\\[5pt]
2305-\frac{30}{7} \beta - \frac{250}{21}\gamma &l=0\\[5pt]
-\frac{41}{2}+\frac{8}{3}\gamma &l=1
\end{array}\right.
\end{align}
As in the twist-four case studied in \cite{Aprile:2017bgs} we note that it is possible to make the anomalous dimensions analytic functions of spin, including for $l=0,1$, by imposing
\be
\beta = 0\,, \qquad \gamma=0\,,
\ee
although we do not have an independent argument for the values of these parameters.

\section{A symmetry of the CFT data}

\label{sec7}

In recent months a wealth of new strong coupling data for double trace operators in $\mathcal{N}=4$ SYM has been computed (see~\cite{Alday:2017xua,Aprile:2017bgs,unmixing} as well as above) both at tree level and one loop in supergravity. We here make the observation that all this new data possesses a nontrivial $Z_2$ symmetry. As we have seen above, the double trace operators in question form natural families. A single family consists of all operators with fixed naive twist, $\Delta_0-l$, fixed $SU(4)$ quantum numbers, and fixed label $i$ distinguishing operators with identical naive quantum  numbers. The spin $l$ is allowed to vary within the family, although it is also useful to separate even and odd spin cases into different families. The anomalous dimensions and appropriately normalised OPE coefficients of the family of operators, are then given as an analytic function of the spin $l$. The statement of the $Z_2$ symmetry is then that under the map
\begin{align}
\text{\bf symmetry:} \qquad l \ \rightarrow\ -l-T(l;a)-3  \ \label{symmetry}
\end{align}
the data for one family of operators maps onto the data for another (possibly the same) operator. Note that here $T(l;a)=\Delta(l;a)-l$ is the full {\em anomalous} twist of the family of operators transformed into, as a function of $l$. 

We illustrate this with a number of examples. The anomalous dimensions of the operators in the $[n,0,n]$ rep with naive twist$=2t$ and additional label $i=1 \dots t-n-1$ were computed in~\cite{Aprile:2017bgs} and reproduced here in~\eqref{anomn0n}.
The above symmetry~\eqref{symmetry} becomes $l\rightarrow -l-2t-3$ (since we are at leading order in $a$ we only need the naive twist $2t=T(l;0)$ in~\eqref{symmetry}). One can check that the anomalous dimension in~\eqref{anomn0n} transforms as
\begin{align}
\eta^{(1)}_{t,l,n,i} \ \rightarrow\  	\eta^{(1)}_{t,-l-2t-3,n,i}=\eta^{(1)}_{t,l,n,i'} \qquad i'=t-i-n\ .
\end{align}
Under the symmetry the family of operators with labels $t,n,i$ maps to the family with labels $t,n,i'=t-i-n$: the symmetry reverses the list of operators with the same naive quantum numbers.

For the anomalous dimensions of $[0,1,0]$ operators computed above (see~\eqref{anom010}) there are two analytic formulae, one for odd spin operators and one for even spin. The symmetry~\eqref{symmetry} (note that in this case naive twist equals $2t+1$ so the symmetry is $l\rightarrow -l-2t-4$) swaps the formula for even spin into that of odd spin and vice versa as well as reversing the label $i$
\begin{align}
	\eta^{(1),\text{even spin}}_{t,l,i} \ \rightarrow\  	\eta^{(1),\text{even spin}}_{t,-l-2t-4,i}=\eta^{(1),\text{odd spin}}_{t,l,i'} \qquad i'=t-i\ \notag\\
		\eta^{(1),\text{odd spin}}_{t,l,i} \ \rightarrow\  	\eta^{(1),\text{odd spin}}_{t,-l-2t-4,i}=\eta^{(1),\text{even spin}}_{t,l,i'} \qquad i'=t-i\
		\label{sym010}
\end{align}

The symmetry also acts on three point functions (after a  universal factor is taken out). For example, consider the 3-point functions of the long singlet operators in~\eqref{3pnt}
\begin{align}
	\widetilde{c}_{t,l,i}:={\langle \mathcal{O}_{2} \mathcal{O}_{2} K_{t,l,0,i} \rangle^2}/{\frac{(t{+}l{+}1)!^2}{(2t{+}2l{+}2)!}}\ .
\end{align} 
These transform under the symmetry as:
\begin{align}
	\widetilde{c}_{t,l,i} \rightarrow \widetilde{c}_{t,-l-2t-3,i}=\widetilde{c}_{t,l,i'}\qquad i'=t-i\ .
\end{align}

Thus far the examples involved data (anomalous dimensions or OPE coefficients) which were all leading in the coupling $a$ and thus the symmetry~\eqref{symmetry} only involved the bare twist $T(l;0)$. But the symmetry should really be thought of in terms of an expansion in $a$. Remarkably, we find the symmetry remains intact even at one loop where the dimension or twist becomes anomalous. For example consider the one-loop twist 4 singlet anomalous dimensions~\eqref{twist4one-loopanomdims} computed in~\cite{Aprile:2017bgs,Alday:2017xua}. The full anomalous twist of these operators is given by
\begin{align}
	T(l;a)&=4+2a\eta_l^{(1)} +2a^2\eta_l^{(2)}+O(a^3)\notag\\
	\eta_l^{(1)}&=\eta^{(1)}_{t=2,l,n=0,i=1}=-\frac{48}{(l+1)(l+6)}\notag\\
	\eta_l^{(2)}&= \frac{1344 (l-7) (l+14)}{(l-1) (l+1)^2 (l+6)^2 (l+8)} -\frac{2304 (2 l+7)}{(l+1)^3 (l+6)^3}
	\label{anom000}
\end{align}
and the symmetry~\eqref{symmetry}, thus becomes (to the relevant order in the coupling $a$) $l\rightarrow -l -7-2a\eta_l^{(1)}$. Under this transformation, the twist itself  should be invariant. One can check that this is indeed true. 
\begin{align}
T(l;a) &\rightarrow T(-l -7-2a\eta_l^{(1)};a)=4+2a\eta_{{-}l{-}7{-}2a\eta_l^{(1)}}^{(1)} +2a^2\eta_{-l-7}^{(2)}+O(a^3)\notag \\
&=4+2a\eta_{-l-7}^{(1)} + 4a^2 \eta_l^{(1)} \frac{\partial}{\partial l} \eta^{(1)}_{-l-7} +   2a^2\eta_{-l-7}^{(2)}+ O(a^3)\notag\\
&=T(l;a)+O(a^3)\ .
\end{align}
The latter equality arises from the identities
\begin{align}
	\eta_{-l-7}^{(1)}=\eta_{l}^{(1)} \qquad    \eta_{-l-7}^{(2)}=\eta_l^{(2)}-  \frac{\partial}{\partial l} (\eta^{(1)}_{l})^2\ .\label{symorder}
\end{align}
Indeed the first term in the expression for $\eta_l^{(2)}$~(\ref{anom000}c) is symmetric under the lowest order symmetry $l\rightarrow -l-7$. The second term in~(\ref{anom000}c) is antisymmetric under the lowest order symmetry. However it is completely determined from the 1 loop anomalous dimension  by the full symmetry, given by $2\frac{\partial}{\partial l} (\eta^{(1)}_{l})^2$. The full symmetry allows us to predict this antisymmetric part.

As our final example we consider the one loop anomalous dimensions computed in this paper, for the (naive) twist 5, [010] operators. As at tree-level the formulae split into two pieces for even and odd spin respectively~\eqref{anom010},\eqref{anom0101loop}. Nevertheless one can check that the symmetry in the form~\eqref{sym010} extends to the one loop case, swapping the even and odd family. Namely we have
\begin{align}
	T^{\text{even}}(l;a) \rightarrow T^{\text{even}}(-l -T^\text{odd}(l;a);a)=T^{\text{odd}}(l;a)\notag\\
	T^{\text{odd}}(l;a) \rightarrow T^{\text{odd}}(-l -T^\text{even}(l;a);a)=T^{\text{even}}(l;a)
\end{align}
These arise from the following 
order by order relations which can be readily checked 
\begin{alignat}{2}
\eta_{-l-7}^{(1)\text{even}}&=\eta_{l}^{(1)\text{odd}} \qquad    \eta_{-l-7}^{(2)\text{even}}&&=\eta_l^{(2)\text{odd}}-  \frac{\partial}{\partial l} (\eta^{(1)\text{odd}}_{l})^2\notag\\
\eta_{-l-7}^{(1)\text{odd}}&=\eta_{l}^{(1)\text{even}} \qquad    \eta_{-l-7}^{(2)\text{odd}}&&=\eta_l^{(2)\text{even}}-  \frac{\partial}{\partial l} (\eta^{(1)\text{even}}_{l})^2
\ .
\end{alignat}

As we have seen in the above examples the symmetry often transforms quantities for one family of operators to other families. In the special cases where the symmetry leaves the family invariant, this has already been seen in the context of large spin CFT analysis in which it has been observed that anomalous dimensions can be expressed in terms of a certain Casimir $J^2$~\cite{Basso:2006nk,Alday:2015eya}.  To see the equivalence with the above symmetry, first note that if we re-express quantities in terms of a shifted spin, $J$ instead of $l$, where
\begin{align}
	J=l+\frac{T+3}2
\end{align}
then the symmetry~\eqref{symmetry} becomes simply 
\begin{align}
\text{\bf symmetry:} \qquad 	J \rightarrow -J\ .
\end{align}
Thus any quantity which   is invariant under this symmetry will clearly be a function of $J^2$. This is essentially the statement made in previous studies~\cite{Basso:2006nk,Alday:2015eya} except that the closely related quantity $J'^2$ rather than $J^2$ was used, where 
\begin{align}
	J'^2=J^2-1/4 = \left(l+\frac{T+3}2\right)^2-\frac14 =\left(l+\frac{T}2+1\right)\left(l+\frac{T}2+2\right)\ .
\end{align}
However, we stress again, that the symmetry transforms many objects non-trivially.



\section*{Acknowledgements}
FA is partially supported by the ERC-STG grant 637844- HBQFTNCER.
PH acknowledges support from STFC grant ST/P000371/1 and this research was also supported by the Munich Institute for Astro- and Particle Physics (MIAPP) of the DFG cluster of excellence ``Origin and Structure of the Universe". JMD and HP acknowledge support from ERC Consolidator grant 648630 IQFT.

\appendix


\section{SCPW}
\label{Appendix--SCPW}

We briefly review the basics of a superblock expansion in $\mathcal{N}=4$ following~\cite{Doobary:2015gia}

The OPE implies the following decomposition of a four point function, 
\bea
\langle \cO_{p_1}\cO_{p_2}\cO_{p_3}\cO_{p_4}\rangle&=& \mathcal{P}^{\rm (OPE)}_{\,\{p_i\}}
\sum_{\{t,\,l,\,\mathfrak{R}\}} A^{\{p_i\}}_{\,\mathfrak{R}}(t |l)\  \mathbb{S}_{\,\mathfrak{R}}^{\{p_i\}}(t |l)
\label{SCOPE1}
\eea
where $t$ is half twist of the exchanged operator, $t=(\Delta-l)/2$, 
\beq\label{OPEprefactor}
\cP^{\rm (OPE)}_{\,\{p_i\}}=
g_{12}^{d} g_{14}^{p_1-d}g_{24}^{p_2-d}g_{34}^{p_3} \qquad \text{with} \quad p_2 \geq p_1,\ p_4 \geq p_3, \ p_2{-}p_1 \leq p_4{-}p_3\ . 
\eeq
and $\mathbb{S}_{\,\mathfrak{R}}^{\{p_i\}}(t |l)$ are given by the $GL(2|2)$ determinantal formula, 
\begin{align}\label{detform2}
\mathbb{S}^{\{p_i\}}_{\mathfrak{R}} &= 
								\left(\frac{x\bar{x}}{y\bar{y}}\right)^{\frac12(\gamma-p_4+p_4)} 
								F^{\alpha\beta\gamma\ula}\qquad 
								\gamma = p_4-p_3, p_4-p_2+2,\dots, {\rm min}(p_1+p_2,p_3+p_4) \notag\\[.2cm]
F^{\alpha\beta\gamma\ula}\ &=\  (-1)^{p +1 } \frac{ s(x,\bar{x},y,\bar{y}) }{(x-\bar{x})(y-\bar{y})}
\det \left( \begin{array}{cc}
						F^X_{\underline\lambda}&	   R		\\
						K_{\underline \lambda}   &   F^Y
		\end{array}\right)\ , \notag
\end{align}
where $p={\rm min}(\alpha,\beta)$, with (using the notation $x_1=x$ and $x_2=\bar{x}$ and similarly for $y$)
\begin{align}
\alpha&=\tfrac12(\gamma-p_{1}+p_2) \quad \beta=\tfrac12( \gamma +p_{3}-p_{4}) \nonumber \\[.2cm]
(F^X_{\underline \lambda})_{ij}&=\Big(
						    [x_i^{\lambda_j  - j} {}_2F_1( \lambda_j + 1 - j+\alpha,  \lambda_j + 1 - j+\beta;  2 \lambda_j + 2 - 2 j+\gamma;  x_i)] 
																	                     \Big)_{  \substack{1\leq i \leq 2\\ 1\leq j\leq p} }\nonumber\\[5pt]
(F^Y)_{ij}&=\Big(
				(y_j)^{i - 1} {}_2F_1(i -\alpha, i -\beta;  2 i -\gamma; y_j)  
														  \Big)_{\substack{1\leq i\leq p\\1\leq j \leq 2}} \nonumber\\[5pt]
(K_\ula)_{ij}&=\Big( 
				-\delta_{i;\,j{-}\lambda_j}  \Big)_{\substack{1\leq i\leq p\\ 1\leq j  \leq p}} \nonumber\\[5pt]
R&=\left(
\begin{array}{cc}
		\frac1{x-y}&\frac1{x-\bar{y}}\\
		\frac1{\bar{x}-y}&\frac1{\bar{x}-\bar{y}}
\end{array}\right)\nonumber\\[5pt]
s(x,\bar{x},y,\bar{y})&
	=(x-y)(x-\bar{y})(\bar{x}-y)(\bar{x}-\bar{y})  
\end{align}
The notation $[\ldots]$ in $F^X$ indicates that only the regular part should be taken, i.e. if $\lambda_j<j$ 
one has to subtract off the first few terms in the Taylor expansion of the hypergeometrics. 
As written above, the determinantal formula for 
$\mathbb{S}^{\{p_i\}}_{\mathfrak{R}}$ deals with all cases in the table, \\
\begin{align}\label{tableapp}
\begin{array}{|c||c|c|c|c|}
\hline
GL(2|2) \text{ rep }\ula     					&    (\Delta{-}l)/2 			& l     			&  \mathfrak{R} 			  	& \text{multiplet type} \\\hline
[0]                                     					&    \gamma/2          			&0    				& [0,\gamma,0]                            	& \text{half BPS}       \\\hline
\left[1^\mu\right]                                 		 	& \gamma/2             			&0				&   [\mu,\gamma{-}2\mu,\mu]		&\text{quarter BPS} \\ 
\left[\lambda, 1^\mu\right]\ (\lambda\geq 2)   	& \gamma/2             			&\lambda{-}2      	&  [\mu,\gamma{-}2\mu{-}2,\mu] 	&  \text{semi-short}  \\\hline
{ [\lambda_1,\lambda_2,2^{\mu_2},1^{\mu_1}]\  (\lambda_2\geq 2)} 
									& \gamma/2{+}\lambda_2{-}2	&\lambda_1{-}\lambda_2&[\mu_1,\gamma{-}2\mu_1{-}2\mu_2{-}4,\mu_1]&\text{long} \\ \hline
\end{array}
\end{align}
Long multiplets are particularly simple, since the determinantal formula factorize. 
The associated Young tableaux contains a $2\times2$ block, and it is convenient to define $\lambda'_2=\mu_2+2$ and $\lambda'_1=\mu_1+\lambda'_2$, 
i.e. the first and second columns have height $\lambda'_1,\lambda'_2$ respectively, with $\lambda'_1,\lambda'_2 \geq 2$.
The explicit expression of the superblock is
\begin{align}\label{flong}
	F_{\text{long}}^{\alpha\beta\gamma\ula}&= 
				(-1)^{\lambda'_1+\lambda'_2}s(x,\bar{x},y,\bar{y})
				\times\frac{F_{\lambda _1}^{\alpha \beta \gamma }\left(x\right) F_{\lambda _2-1}^{\alpha \beta \gamma}\left(\bar{x}\right)- (x \leftrightarrow \bar{x})}{x-\bar{x}}\notag\\
				&\rule{4.4cm}{0pt} \times\frac{G_{\lambda' _1}^{\alpha \beta \gamma }\left(y\right) G_{\lambda' _2-1}^{\alpha \beta\gamma }\left(\bar{y}\right)-(y\leftrightarrow \bar{y})}{y-\bar{y}}\\[.2cm]
	F^{\alpha\beta\gamma}_\lambda(x)& :=
	x^{\lambda-1}{}_2F_1(\lambda+\alpha,\lambda+\beta;2\lambda+\gamma;x)\notag\\
	G^{\alpha\beta\gamma}_{\lambda'}(y)&:=
	y^{\lambda'-1}{}_2F_1(\lambda'-\alpha,\lambda'-\beta;2\lambda'-\gamma;y)\ .
\end{align}
Alternatively we can rewrite $\mathbb{S}_{\,\mathfrak{R};\, {\rm long} }^{\{p_i\}}\rightarrow \LL$ in terms of 
conformal blocks  $\CB^{\,t|l}$ and $SU(4)$ harmonics $Y_{nm}$ 
commonly introduced in the literature, 
\bea
&& \label{longmultiplet}
\LL^{\{p_i\}}_{\,nm}(t|l)=\frac{s(x,\bar{x},y,\bar{y})}{  \rule{0pt}{.45cm}  (y\bar{y})^2}\, 
\frac{ \CB^{\,2+t|l} }{   \rule{0pt}{.45cm} u^{2+ \frac{ p_{43}}{2}  } } 
\frac{(n+1)!m!}{ \rule{0pt}{.45cm} (n+2+ p_{43} )_{n+1} (m+1+ p_{43} )_m }Y_{nm}\ .
\eea
where 
\bea
\label{Confblock}
\CB^{\,t|l}&=&(-1)^l\ 
				 \frac{x^{t+l+1}\, \bar{x}^{t}\ \bF_{t+l }(x)\bF_{t-1}(\bar{x})- \bar{x}^{t+l+1}\, x^{t}\ \bF_{t-1}(x)\bF_{t+l }(\bar{x}) }{x-\bar{x}}\\[.2cm]
 \label{SU4harm}
Y_{nm}&=&	
			 - \frac{ \bP_{n+1}(y) \bP_m(\bar{y})-  \bP_m(y) \bP_{n+1}(\bar{y})  }{y-\bar{y}} \\[.2cm]
\bF_{t}(x)&=&
			\,_2F_1\left( t- \tfrac{p_{12} }{2},t+\tfrac{p_{34}}{2},2t;x\right),  \\[.2cm]
\bP_{n}(y)&=&
			\, {y}\, {\rm JP}^{(p_1-d_{12}|p_2-d_{12})}_n\left(\frac{2}{y}-1 \right)\ .
\eea
The corresponding $SU(4)$ representation translates to $[n-m, 2m+p_{43} ,n-m]$, with
$m=p_{34}/2+\gamma/2-\lambda'_1$, and $n=p_{34}/2+\gamma/2-\lambda'_2$.


\section{On supergravity correlators }
\label{Appendix--SugraCorrelators}


\subsection{Anatomy of $\Db$ functions}
\label{Appendix--AnatomyOfDbarfunctions}

Any $\Dbar_{\delta_1\delta_2\delta_3\delta_4}$ with integer $\sigma\equiv(\delta_1+\delta_2-\delta_3-\delta_4)/2\ge 0$ can be written very explicitly as 
\beq\label{DbaruY}
\Dbar_{\delta_1\delta_2\delta_3\delta_4}= u^{-\sigma}\, \Dbar_{\delta_1\delta_2\delta_3\delta_4}^{\rm\, sing} +  \Dbar_{\delta_1\delta_2\delta_3\delta_4}^{\rm\, analytic}+\,   \log(u)\,\Dbar_{\delta_1\delta_2\delta_3\delta_4}^{\rm\, log}\ , 
\eeq
The first two functions are, 
\bea\label{Dsingularseries}
\Dbar_{\delta_1\delta_2\delta_3\delta_4}^{\rm\,sing}&=& 
									\sum_{n=0}^{\sigma-1} \frac{(-u)^{n}}{n!}\ 
									\Gamma[\sigma-n]\ \Lambda^{\delta_3\delta_4}_{\delta_1-\sigma\delta_2-\sigma}(n)\ {F}^{\,\delta_2-\sigma+n | \delta_3+n}_{\, \delta_3+\delta_4+2n}(1-v)\ ,\\
 \label{Dlogseries}
\Dbar_{\delta_1\delta_2\delta_3\delta_4}^{\rm\, log}&=& 
									(-)^{\sigma+1}  \sum_{n=0}^\infty \frac{u^n}{n!(\sigma+n)!}\  
									\Lambda^{\delta_1\delta_2}_{\delta_3+\sigma\delta_4+\sigma}(n)\  {F}^{\, \delta_2+n |\delta_3+\sigma+n}_{\, \delta_1+\delta_2+2n}(1-v)\ ,
\eea
where we defined\footnote{ $\Dbar_{\delta_1\delta_2\delta_3\delta_4}^{\rm\, sing}=0$ when $\sigma=0$.}
\begin{align}
F^{a|b}_c(x)\equiv\,_2 F_1[a,b;c](x),
& &
\Lambda^{\delta_1\delta_2}_{\delta_3\delta_4}(n)\equiv \frac{ \Gamma[\delta_1+n]\Gamma[\delta_2+n]\Gamma[\delta_3+n]\Gamma[\delta_4+n] }{\Gamma[\delta_1+\delta_2+2n]}\ .
\end{align}
The expression for $\Dbar_{\delta_1\delta_2\delta_3\delta_4}^{\rm\, analytic}$ is given by
\begin{align}
	\Dbar_{\delta_1\delta_2\delta_3\delta_4}^{\rm analytic}= 
	(-)^\sigma \sum_{n,m\ge 0} \frac{u^n }{n!(\sigma+n)!} \Lambda^{\delta_1\delta_2}_{\delta_3+\sigma\delta_4+\sigma}(n) 
				\frac{ (\delta_2+n)_m (\delta_3+\sigma+n)_m }{ (\delta_1+\delta_2+2n)_m }\,\mathfrak{f}_{nm} \frac{(1-v)^m}{m!\ }
\end{align}
where 
\bea
\mathfrak{f}_{nm}
		&= &\Big[ +\psi(n+1)+\psi(\sigma+1+n) + 2 \psi(\delta_1+\delta_2+2n+m) \rule{0cm}{.6cm} \nonumber\\
& & \rule{.5cm}{0cm} -\psi(\delta_4+\sigma+n)-\psi(\delta_1+n) -\psi(\delta_3+\sigma+n+m) -\psi(\delta_2+n+m)\Big]\,  \nonumber
\eea
The representation \eqref{DbaruY} is very useful in an OPE expansion. On the other hand, any $\Dbar_{\delta_1\delta_2\delta_3\delta_4}$ can be obtained as operators acting on $\Db_{1111}$. The set of operators is
\begin{align}
\Dbar_{\delta_1+1,\delta_2+1,\delta_3,\delta_4}&=-\du \Dbar_{\delta_1\delta_2\delta_3\delta_4},\nonumber\\
\Dbar_{\delta_1,\delta_2,\delta_3+1,\delta_4+1}&=(\delta_3+\delta_4-\Sigma-u\du)\Dbar_{\delta_1\delta_2\delta_3\delta_4},\nonumber\\
\Dbar_{\delta_1,\delta_2+1,\delta_3+1,\delta_4}&=-\dv\Dbar_{\delta_1\delta_2\delta_3\delta_4},\nonumber\\
\Dbar_{\delta_1+1,\delta_2,\delta_3,\delta_4+1}&=(\delta_1+\delta_4-\Sigma-v\dv)\Dbar_{\delta_1\delta_2\delta_3\delta_4},\nonumber\\
\Dbar_{\delta_1,\delta_2+1,\delta_3,\delta_4+1}&=(\delta_2+u\du+v\dv)\Dbar_{\delta_1\delta_2\delta_3\delta_4},\nonumber\\
\Dbar_{\delta_1+1,\delta_2,\delta_3+1,\delta_4}&=(\Sigma-\delta_4+u\du+v\dv)\Dbar_{\delta_1\delta_2\delta_3\delta_4},
\end{align}
 Defining $\Dbar_{\delta_1\delta_2\delta_3\delta_4}$ in this way provides a resummation of the series expansions in \eqref{DbaruY}. In fact, 
$\Db_{1111}$ admits the following representation in term of polylogarithms \cite{Usyukina:1992jd}
\beq\label{one-loop-box}
\Dbar_{1111}=-\log(u)\,\frac{{\rm Li}_1(x)-{\rm Li}_1(\bar{x})}{x-\bar{x} }+2\, \frac{{\rm Li}_2(x)-{\rm Li}_2(\bar{x})}{x-\bar{x}}\ ,
\eeq
where $u=x \bar{x}$ and $v=(1-x)(1-\bar{x})$.


\subsection{Examples of supergravity correlators}
\label{Appendix--ExamplesSugra}

In \cite{unmixing} section 2.2, an algorithm was given to provide a $\Db_{\delta_1\delta_2\delta_3\delta_4}$ representation of the supergravity amplitude of Rastelli and Zhou \cite{Rastelli:2016nze}. 
Following the same implementation it is simple to get results for the family of correlators $\langle \cO_p\cO_{p+1}\cO_q\cO_{q+1}\rangle$. We list the first few cases here below:  
\bea
\cH_{p,p+1,2,3}=& 
									u^p         & 	\Db_{p,p+3,2,3}  \label{corr_23} \\[.2cm]								
\cH_{p,p+1,3,4}=& 
									u^p \Big[ &	\frac{1}{2}\sigma \Db_{p-1,p+3,2,4}+ \tau \Db_{p-1,p+3,3,3} + \nonumber\\
									&	       & 	\frac{1}{p-2} \Db_{p,p+3, 2, 3} + \left( \frac{1}{p-2}+\frac{1}{2}\sigma +\tau \right)   \Db_{p, p+3, 3, 4}\Big] \label{corr_34}\\[.2cm]							
\cH_{p,p+1,4,5}=& 
									u^p \Big[ & 	\sigma\tau \left(\Db_{p-2,q+3,3,4}+\Db_{p-1,q+3,3,5}+\Db_{p-1,p+3,4,4}+\Db_{p,p+3,4,5}\right) +\nonumber\\
									& 	      &         \frac{1}{6} \sigma^2 \left( 2\Db_{p-2,p+3,2,5}+ 2 \Db_{p-1,p+3,3,5}+\Db_{p,p+3,4,5} \right)+\nonumber \\
									&	      &		\frac{1}{2} \tau^2 \left( 2\Db_{p-2,p+3,4,3}+ 2 \Db_{p-1,p+3,4,4}+\Db_{p,p+3,4,5} \right)+\nonumber\\
									&	      & 	\frac{1}{p-3}\sigma \left(\Db_{p-1,p+3,2,4}+ \Db_{p-1,p+3,3,5}+\Db_{p,p+3,3,4}+\Db_{p,p+3,4,5} \right)+\nonumber\\
									&	      &		\frac{2}{p-3}\tau \left( \Db_{p-1,p+3,3,3}+\Db_{p-1,p+3,4,4}+\Db_{p,p+3,3,4}+\Db_{p,p+3,4,5} \right) +\nonumber\\
									&            &		\frac{1}{(p-2)(p-3)} \left( 2 \Db_{p,p+3,2,3} + 2 \Db_{p,p+3,3,4}+\Db_{p,p+3,4,5} \right) \Big] \label{corr_45}		
\eea
and finally
\bea
\cH_{p,p+1,5,6}=& u^p \Big[ & \sum_{k=0}^3 \frac{\sigma^3}{4k!} \Db_{p-3+k,p+3,2+k,6} +\frac{\tau^3}{k!} \Db_{p-3+k,p+3,5,3+k} +  \nonumber\\
&&
\sum_{k=0}^2\sum_{m=0}^1    \frac{3\tau^2\sigma }{2k!} \Db_{q-3+k+m,q+3,4+m,4+k}  
										        		+  \frac{\sigma^2\tau}{k!} \Db_{q-3+k+m,q+3,3+k,5+m} + \nonumber\\
&&
\sum_{k=0}^2\sum_{m=0}^1 \frac{\sigma^2}{(p-4)k!} \Db_{p-2+k,p+3,2+k+m,5+m} + \frac{3\tau^2}{(p-4)k!}  \Db_{p-2+k,p+3,4+m,3+k+m}+\nonumber\\
&& 
\frac{3\sigma\tau}{q - 4} \sum_{k=0}^1\sum_{m=0}^1 (\Dbar_{p - 2 + k, p + 3, 3+m, 4 + m + k}+\Dbar_{p - 1 + k, p + 3, 4 + m, 4 + m + k})+\nonumber\\
&&
\frac{3\sigma}{(p-4)(p-3)}\sum_{k=0}^1\sum_{m=0}^2 \frac{1}{m!} \Db_{p-1+k,p+3,2+m+k,4+m}+\nonumber\\ 
&&
\frac{6\tau}{(p-4)(p-3)} \sum_{k=0}^1\sum_{m=0}^2 \frac{1}{m!} \Db_{p-1+k,p+3,3+m,3+k+m}+\nonumber\\
&&
\frac{6}{(p-4)(p-3)(p-2)} \sum_{k=0}^3 \frac{1}{k!} \Db_{p,p+3,2+k,3+k}	\Big]	 \label{corr_56}					
\eea
The expression of the supergravity amplitude in Mellin space assumes an ordering of the charges. 
Thus in the last step of the algorithm we obtained the correlators of interest by acting with symmetries. 
We also used the reflection property $\Db_{\delta_1,\delta_2,\delta_3,\delta_4}=\Db_{\Sigma-\delta_1,\Sigma-\delta_2,\Sigma-\delta_3,\Sigma-\delta_4}$ 
where $\Sigma=\frac{\delta_1+\delta_2+\delta_3+\delta_4}{2}$. Other identities among $\Db_{\delta_1\delta_2\delta_3\delta_4}$ 
can be used to represent the final result in equivalent ways. For example, in the proof of \eqref{proof_norma} we considered the identity, 
$\overline{D}_{\delta_1\delta_2\delta_3\delta_4}(u,v)=v^{\delta_1+\delta_4-\Sigma}\, \overline{D}_{\delta_2\delta_1\delta_4\delta_3}(u,v)$.


\begin{thebibliography}{99}
	
	
	\bibitem{1}
	J.~Maldacena, {\it Adv.~Theor.~Math.~Phys.} {\bf 2} (1998) 231
	[arXiv:hep-th/9711200]
	
	\bibitem{2}
	S.~Gubser, I.~Klebanov and A.~Polyakov, {\it Phys.~Lett.} {\bf B428}
	(1998) 105 [arXiv:hep-th/9802109]
	
	\bibitem{3}  
	E.~Witten, {\it Adv.~Theor.~Math.~Phys.} {\bf 2} (1998) 253
	[arXiv:hep-th/9802150].
	
	
	\bibitem{Liu:1998ty}
	H.~Liu and A.~A.~Tseytlin,
	Phys.\ Rev.\ D {\bf 59} (1999) 086002
	doi:10.1103/PhysRevD.59.086002
	[hep-th/9807097].
	
	\bibitem{Arutyunov:1999fb} 
	G.~Arutyunov and S.~Frolov,
	Nucl.\ Phys.\ B {\bf 579}, 117 (2000)
	doi:10.1016/S0550-3213(00)00210-8
	[hep-th/9912210].
	
	
	\bibitem{Arutyunov:2002fh}
	G.~Arutyunov, F.~A.~Dolan, H.~Osborn and E.~Sokatchev,
	Nucl.\ Phys.\ B {\bf 665} (2003) 273
	[hep-th/0212116].
	
	\bibitem{Arutyunov:2003ae}
	G.~Arutyunov and E.~Sokatchev,
	Nucl.\ Phys.\ B {\bf 663} (2003) 163
	[hep-th/0301058].
	
	
	\bibitem{Dolan:2006ec} 
	F.~A.~Dolan, M.~Nirschl and H.~Osborn,
	Nucl.\ Phys.\ B {\bf 749}, 109 (2006)
	doi:10.1016/j.nuclphysb.2006.05.009
	[hep-th/0601148].
	
	\bibitem{Berdichevsky:2007xd}
	L.~Berdichevsky and P.~Naaijkens,
	JHEP {\bf 0801} (2008) 071
	[arXiv:0709.1365 [hep-th]].
	
	\bibitem{Uruchurtu:2008kp} 
	L.~I.~Uruchurtu,
	JHEP {\bf 0903}, 133 (2009)
	doi:10.1088/1126-6708/2009/03/133
	[arXiv:0811.2320 [hep-th]].
	
	
	
	\bibitem{Uruchurtu:2011wh}
	L.~I.~Uruchurtu,
	JHEP {\bf 1108} (2011) 133
	[arXiv:1106.0630 [hep-th]].
	
		\bibitem{Rastelli:2016nze} 
	L.~Rastelli and X.~Zhou,
	Phys.\ Rev.\ Lett.\  {\bf 118}, no. 9, 091602 (2017)
	doi:10.1103/PhysRevLett.118.091602
	[arXiv:1608.06624 [hep-th]].
	
	
	\bibitem{Cardona:2017tsw}
	C.~Cardona,
	arXiv:1708.06339 [hep-th].
	
	\bibitem{Giombi:2017hpr}
	S.~Giombi, C.~Sleight and M.~Taronna,
	arXiv:1708.08404 [hep-th].
	
	
	\bibitem{1612.03891}
	O.~Aharony, L.~F.~Alday, A.~Bissi and E.~Perlmutter,
	arXiv:1612.03891 [hep-th].
	
	\bibitem{Alday:2017xua}
	L.~F.~Alday and A.~Bissi,
	arXiv:1706.02388 [hep-th].
	
	
	\bibitem{Aprile:2017bgs}
	F.~Aprile, J.~M.~Drummond, P.~Heslop and H.~Paul,
	arXiv:1706.02822 [hep-th].
	



\bibitem{unmixing} 
  F.~Aprile, J.~M.~Drummond, P.~Heslop and H.~Paul,
  arXiv:1706.08456 [hep-th].
	
	
\bibitem{Alday:2017vkk}
L.~F.~Alday and S.~Caron-Huot,
arXiv:1711.02031 [hep-th].
	
	
	\bibitem{Eden:2000bk}
	B.~Eden, A.~C.~Petkou, C.~Schubert and E.~Sokatchev,
	Nucl.\ Phys.\ B {\bf 607} (2001) 191
	[hep-th/0009106].
	
	\bibitem{Dolan:2000ut}
	F.~A.~Dolan and H.~Osborn,
	Nucl.\ Phys.\ B {\bf 599} (2001) 459
	[hep-th/0011040].
	
	\bibitem{Dolan:2001tt}
	F.~A.~Dolan and H.~Osborn,
	Nucl.\ Phys.\ B {\bf 629} (2002) 3
	doi:10.1016/S0550-3213(02)00096-2
	[hep-th/0112251].
	
	\bibitem{Dolan:2004iy}
	F.~A.~Dolan and H.~Osborn,
	Annals Phys.\  {\bf 321} (2006) 581
	[hep-th/0412335].
	
	\bibitem{Doobary:2015gia}
	R.~Doobary and P.~Heslop,
	JHEP {\bf 1512} (2015) 159
	doi:10.1007/JHEP12(2015)159
	[arXiv:1508.03611 [hep-th]].
	
	
	
	
	\bibitem{hep-th/9903196}
	E.~D'Hoker, D.~Z.~Freedman, S.~D.~Mathur, A.~Matusis and L.~Rastelli,
	Nucl.\ Phys.\ B {\bf 562} (1999) 353
	doi:10.1016/S0550-3213(99)00525-8
	[hep-th/9903196].
	
	\bibitem{Arutyunov:2000py} 
	G.~Arutyunov and S.~Frolov,
	Phys.\ Rev.\ D {\bf 62}, 064016 (2000)
	doi:10.1103/PhysRevD.62.064016
	[hep-th/0002170].
	
	\bibitem{Alday:2016njk}
	L.~F.~Alday,
	arXiv:1611.01500 [hep-th].
	
	\bibitem{0907.0151}
	I.~Heemskerk, J.~Penedones, J.~Polchinski and J.~Sully,
	JHEP {\bf 0910} (2009) 079
	doi:10.1088/1126-6708/2009/10/079
	[arXiv:0907.0151 [hep-th]].
	
	\bibitem{Fitzpatrick:2011dm}
	A.~L.~Fitzpatrick and J.~Kaplan,
	JHEP {\bf 1210} (2012) 032
	doi:10.1007/JHEP10(2012)032
	[arXiv:1112.4845 [hep-th]].
	
	\bibitem{Usyukina:1992jd}
	N.~I.~Usyukina and A.~I.~Davydychev,
	Phys.\ Lett.\ B {\bf 298} (1993) 363.
	doi:10.1016/0370-2693(93)91834-A
	
	\bibitem{Basso:2006nk}
	B.~Basso and G.~P.~Korchemsky,
	Nucl.\ Phys.\ B {\bf 775} (2007) 1
	doi:10.1016/j.nuclphysb.2007.03.044
	[hep-th/0612247].
	
	
	
	
	
	\bibitem{Alday:2015eya}
	L.~F.~Alday, A.~Bissi and T.~Lukowski,
	JHEP {\bf 1511} (2015) 101
	doi:10.1007/JHEP11(2015)101
	[arXiv:1502.07707 [hep-th]].
	
	
	
	
	
	
	
	
	
	
	
	

\end{thebibliography}
\end{document}